\documentclass[12pt]{article}
\usepackage{color}
\usepackage{ulem}

\usepackage{graphicx,amsmath,amsfonts,amssymb,amsthm,bbm,mathrsfs,times}

\date{29-JAN-2014}

\theoremstyle{plain}
\newtheorem{satz}{Theorem}[section]
\newtheorem{lemma}[satz]{Lemma}
\newtheorem{koro}[satz]{Corollary}
\theoremstyle{definition}

\newtheorem{bemerkung}[satz]{Remark}

\newcommand{\cO}{{\cal O}}

\newcommand{\cP}{{\cal P}}

\renewcommand{\d}{{\rm d}}
\newcommand{\fer}[1]{(\ref{#1})}
\newcommand{\scalprod}[2]{\left\langle {#1}, {#2}\right\rangle}
\newcommand{\R}{\mathcal{R}}
\renewcommand{\S}{\mathrm{s}}
\newcommand{\func}[1]{{\rm #1}}
\newcommand{\e}{{\rm e}}
\renewcommand{\frak}{\mathfrak}

\newcommand{\change}
{{\marginpar{\#}}}        
\newcommand{\eps}{{\varepsilon}}        
                
\newcommand{\vth}{\vartheta}

\newcommand{\one}{{\bf 1}}

\newcommand{\cG}{{\mathcal G}}

\newcommand{\hV}{\widehat{V}}

\newcommand{\cirS}{\mathop{\bigcirc\kern -.73em {\scriptstyle{\rm S}}}}

\begin{document}

\title{Suppression of Decoherence \\
 by Periodic Forcing}

\author{
Volker Bach\footnote{Institut f\"ur Analysis und Algebra, TU~Braunschweig, 
38106 Braunschweig, Germany; \texttt{v.bach@tu-bs.de}}
\and 
Walter de Siqueira Pedra\footnote{Universidade de Sao Paulo, Instituto 
de Fisica - Departamento de Fisica Matematica, Caixa Postal 66318,
05314-970 Sao Paulo/SP, Brasil; \texttt{wpedra@if.usp.br}}
\and
Marco Merkli\footnote{Department of Mathematics and Statistics, Memorial 
University of Newfoundland, St. John's, Newfoundland, Canada A1C 5S7;
\texttt{merkli@mun.ca}; Partially supported by NSERC}
\and
Israel Michael Sigal\footnote{Department of Mathematics, University of 
Toronto, Toronto, ON M5S 2E4, Canada; \texttt{im.sigal@utoronto.ca}}
}

\maketitle

\noindent
\begin{center}
\begin{tabular}{c}
\textit{Dedicated to Herbert Spohn,}\\ 
\textit{for his scientific contributions and his friendship.}
\end{tabular}
\end{center}

\begin{abstract}
We consider a finite-dimensional quantum system coupled to a thermal
reservoir and subject to a time-periodic, energy conserving
forcing. We show that%
\color{black}, if a certain \textit{dynamical
  decoupling condition} is fulfilled, then \color{black}
the periodic forcing counteracts the decoherence induced by the
reservoir: for small system-reservoir coupling $\lambda$ and small
forcing period $T$, the system dynamics is approximated by an energy
conserving and non-dissipative dynamics, which preserves
coherences. For times up to order $(\lambda T)^{-1}$, the difference
between the true and approximated dynamics is of size $\lambda
+T$. Our approach is rigorous and combines Floquet and spectral
deformation theory. We illustrate our results on the spin-fermion
model and recover previously known, heuristically obtained results.
\end{abstract}

\newpage
\section{Introduction and main results}

The phenomenon of decoherence -- the destruction of quantum coherence -- which leads to the transition from quantum to classical behaviour, is one of the central phenomena in quantum physics. Decoherence is crucial for processing quantum information and it presents the main obstacle to building quantum computers. 

The most relevant aspect for quantum information processing is decoherence due to the interaction with the environment \cite{jzk..,PSE}, and its mathematical understanding is at the heart of possible schemes to prevent it, or to slow it down. The present work investigates the control of environment-induced decoherence by subjecting the system to a periodic external force. Periodic forcing as a means (among others) to dampen decoherence has been proposed in the recent theoretical physics literature \cite{Facchi, Vio}. Our contribution to the topic in the present paper is a mathematical analysis of the evolution of periodically driven open quantum systems and a rigorous proof that decoherence is suppressed by periodic forcing%
\color{black}, provided the latter fulfills the \textit{Dynamical Decoupling Condition}~\eqref{dyn.cond}, which we adopt from \cite{Facchi, Vio}.\color{black}

Let $\rho_{\rm s}(t)$ be the \textit{reduced density matrix} of a quantum system, not subject to external forcing, but in contact with a thermal reservoir at temperature $1/\beta>0$. Let $H_\S$ be the Hamiltonian of the system, with eigenvalues $E_i$ and orthonormalized eigenvectors $\varphi_i$. The (energy basis) matrix elements of $\rho_\S(t)$ are given by $[\rho_\S(t)]_{i,j}=\scalprod{\varphi_i}{\rho_{\rm s}(t)\varphi_j}$. If the system does not interact with the reservoir then its dynamics is unitary and simply given by 
$[\rho_\S(t)]_{i,j}=\mathrm{e}^{i t(E_i-E_j)}[\rho_\S(0)]_{i,j}$. When the interaction is turned on, however, the system dynamics becomes irreversible, as energy transferred to the reservoir is dissipated into the vastness of its (infinite) volume. In generic situations where energy exchange takes place, the system plus its environment evolve into their \textit{joint} equilibrium state at temperature $1/\beta$, regardless of the initial state of the system alone. This dynamical process is called thermalization. The reduction of the final equilibrium state to the system alone (obtained by tracing out the reservoir degrees of freedom) is, to lowest approximation in the interaction, the Gibbs equilibrium state
 $\rho_{\rm Gibbs} =\mathrm{e}^{-\beta H_\S}/{\rm Tr}\mathrm{e}^{-\beta H_\S}$. This means that 
\color{black}
\begin{align*}
\lim_{t\rightarrow\infty} \rho_{\rm s}(t) \ = \ \rho_{\rm Gibbs} + R,
\end{align*}
\color{black}
where $R$ is a remainder of the order of the system-reservoir interaction strength. The approximate final state $\rho_{\rm Gibbs}$ is \textit{diagonal in the energy basis}. Since off-diagonal density matrix elements represent quantum coherences \cite{jzk..}, this means that thermalization causes decoherence. Open systems with conserved energy $H_\S$ are also widely studied (they are often explicitly solvable), and one can show that typically, off-diagonal reduced density matrix elements decay to zero, while the diagonal elements are time-independent (here, thermalization does not occur) \cite{PSE}. The latter process is called ``pure dephasing'' and is a form of decoherence. It thus appears that a generic effect of noise is to drive off-diagonal system density matrix elements to zero (modulo some small error). In \cite{dec1,dec2,dec3}, thermalization and decoherence are analyzed rigorously. It is shown there that the decay of off-diagonals is exponentially quick in time, with decay rates given by imaginary parts of complex system energies, associated to quantum resonances.

What happens if we superpose, to the noise coupling, a \textit{structured}
(time-periodic) external force $H_{\mathrm{c}}(t)$ acting on the system? Let 
$T$ be the period of the forcing, and let $\lambda $ be the
system-environment coupling strength. Both $T$ and $\lambda $ are taken to
be small. We show in Corollary~\ref{decocoro} below that for times up to
order $(\lambda T)^{-1}$, the dynamics is given by 
\color{black}
\begin{align*}
[\rho_\S(t)]_{i,j}
\ = \ 
\mathrm{e}^{i\Phi _{i,j}(t)}[\rho_\S(0)]_{i,j} + \mathcal{O}(\lambda +T),
\end{align*}
\color{black}
where $\Phi _{i,j}(t)$ is a real phase. The off-diagonal density matrix
elements do \textit{not} decay, so the forcing has counteracted the
decoherence effect of the reservoir.
\color{black} Comparing to the usual time asymptotics for matrix
elements which generically decay on the scale $t \sim \lambda^{-2}$,
the forcing has a noticeable effect provided $T \ll |\lambda|$ (see
Remark~\ref{rem-1.0,1}).\color{black}
This result is
a consequence of Theorem \ref{mainthm} below, which is our main result. In
this theorem, we prove that if $\lambda$ and $T$ are small enough, then the system dynamics is close to the one generated by the system Hamiltonian \textit{plus the forcing term}, but \textit{without the interaction with the heat bath}. The approximating effective dynamics has thus the coherence-preserving property mentioned above, since it is assumed that the forcing operator commutes with the system Hamiltonian.

It is not difficult to see that, in general, if the interaction of the system with the reservoir commutes with $H_{\mathrm{c}}$, then reservoir-induced, pure-phase decoherence cannot be suppressed by the additional periodic force on the system. We must
therefore impose a condition of \textquotedblleft
effective\textquotedblright\ coupling of the external force to the
system-reservoir interaction, called the \textit{dynamical decoupling condition%
} (\ref{dyn.cond}), which has been identified in the literature before \cite%
{Facchi,Vio}. To arrive at our results, we link the reduced dynamics to the
spectrum of an effective propagator and employ spectral deformation
techniques to analyze the latter.

\medskip
{\bf Model and main results.\ } 
\color{black} We consider a small system,
$\mathcal{S}$, interacting with the environment, also called the
\textit{reservoir}, $\mathcal{R}$, described by a free quantum
field. The main application we have in mind is a small system, such as
one or a few qubits, which is coupled to the quantized radiation field
or to the phonon field - both being boson fields.

Although free Bose fields are natural for modelling environments, in
order to keep the exposition technically as simple as possible we
assume that the reservoir $\mathcal{R}$ consists of free fermions. In the
fermionic case, the interaction is a bounded operator and is easier to
deal with. We expect, however, that, using the results and methods of
\cite{JP1, BFS, dec2, dec3}, where such models with bosonic reservoirs were
analyzed rigorously, our treatment can be extended to the bosonic case
without altering the essence of our conclusions and of our approach
(a further discussion is given in Remark~\ref{rem-1.5}).
\color{black}

The forced dynamics of the system is generated by a time-dependent
Hamiltonian 
\[
H_{\mathrm{s}}+H_{\mathrm{c}}(t)
\]%
acting on ${\mathbb{C}}^{d}$, the pure state Hilbert space of the system  $\mathcal{S}$. $H_{\mathrm{s}}$ is the
intrinsic Hamiltonian of $\mathcal{S}$. $H_{\mathrm{c}}(t)$ is an external forcing, the ``control term''. We assume that \textit{the control term has period $T$ and commutes with the system Hamiltonian at all times $t\geq 0$ \color{black}(see Remark~\ref{rem-1.6})\color{black}}:
\begin{eqnarray}
H_{\mathrm{c}}(t+T) &=&H_{\mathrm{c}}(t),
\label{perc} 
\\ 
\label{force-commute}
{}[H_{\mathrm{s}},H_{\mathrm{c}}(t)] &=&0.
\end{eqnarray}
The commutator $\delta_{\mathrm{s,c}}=\delta
_{\mathrm{s}}+\delta_{\mathrm{c}}(t)=i[H_{\mathrm{s}},\,\cdot
  \,]+i[H_{\mathrm{c}}(t),\,\cdot \,]$ is a symmetric derivation on
the bounded linear operators $\mathcal{B}({\mathbb{C}}^{d})$. It
determines the time-dependent Heisenberg evolution
$\tau_{t,0}^{\mathrm{s,c}}$ of the forced system,

\begin{equation}
\partial _{t}\tau _{t,0}^{\mathrm{s,c}}(A)=\delta _{{\mathrm{s,c}}%
}(t)\big(\tau _{t,0}^{\mathrm{s,c}}(A)\big),\qquad \tau _{0,0}^{\mathrm{s,c}}(A)=A.
\label{m101}
\end{equation}
The map $A\mapsto \tau _{t,0}^{\mathrm{s,c}}(A)$ is a $\ast $--automorphism on ${\cal B}({\mathbb C}^d)$. We have the representation 
\begin{equation}
\tau _{t,0}^{\mathrm{s,c}}(A)= V_{\mathrm{c}}(t)\mathrm{e}^{\mathrm{i}tH_{\mathrm{s}}} A
\mathrm{e}^{-\mathrm{i}tH_{\mathrm{s}}}V_{\mathrm{c}}(t)^{\ast},
\label{scdyn}
\end{equation}
where the unitary 
$V_{\mathrm{c}}(t)$ is given by
\begin{equation}
V_{\mathrm{c}}(t)=\mathrm{id}_{\mathbb{C}^{d}}+\sum_{n\geq
1}i^{n}\int_{0<t_{1}<\ldots <t_{n}<t}\,H_{\mathrm{c}}(t_{n})\ldots H_{%
\mathrm{c}}(t_{1})\mathrm{d}t_{1}\ldots \mathrm{d}t_{n}.  \label{vc}
\end{equation}

Consider now the reservoir $\mathcal{R}$. 
Its observable
algebra is the canonical anti-commutation relation (CAR) $\ C^{\ast }$%
--algebra ${\mathcal{V}}_{\mathcal{R}}$, generated by the annihilation and
creation operators $a(f)$, $a^{\ast }(f)$, $f\in L^{2}(\mathbb{R}^{3})$,
acting on the antisymmetric Fock space over the one-particle space $L^2({\mathbb R}^3)$. The creation and annihilation operators fulfill the CAR 
\[
a(f_{1})a^{\ast }(f_{2})+a^{\ast }(f_{2})a(f_{1}) =\langle
f_{1},f_{2}\rangle , \quad 
a(f_{1})a(f_{2})+a(f_{2})a(f_{1}) =0.
\]
The reservoir dynamics is the Bogoliubov automorphism group
\begin{equation}
\tau _{t}^{{\mathcal{R}}}\left( a(f)\right) =a(\mathrm{e}^{ith_{1}}f),\quad
f\in L^{2}(\mathbb{R}^{3}),\ t\in \mathbb{R}.
\label{definition automorphism reservoir}
\end{equation}%
Here, $h_{1}$ is the diagonalized one--particle Hamiltonian, represented by the 
multiplication operator $f(\mathbf{p})\mapsto |\mathbf{p}|f(\mathbf{p})$ on $%
L^{2}(\mathbb{R}^{3})$. We denote by $\delta _{\mathrm{r}}$ the generator of
the strongly continuous group $\tau _{t}^{{\mathcal{R}}}$, i.e., $%
\partial _{t}\tau _{t}^{\mathcal{R}}(A)=\delta _{\mathrm{r}}(\tau _{t}^{%
\mathcal{R}}(A))$ for all $A\in {\mathcal{V}}_{{\mathcal{R}}}$. The thermal reservoir state $\omega _{\mathcal{R}}$ (the $(\beta ,\tau_t^{\mathcal{R}})$--KMS state at inverse temperature $0<\beta<\infty $) is the quasi-free state satisfying $\omega_{\mathcal R}(a(f)a(g))=\omega_{\mathcal R}(a(f))=0$ and
\begin{equation}
\omega _{\mathcal{R}}\left( a^{\ast }(f)a(g)\right) =\langle g,[1+\mathrm{e}%
^{\beta h_{1}}]^{-1}f\rangle .  \label{gibbsreservoir}
\end{equation}

The observable algebra of the total system is the $C^{\ast }$--algebra ${\mathcal{V}}:=\mathcal{B}(\mathbb{C}^{d})\otimes 
{\mathcal{V}}_{\mathcal{R}}$. The total dynamics is  the solution of the differential equation 
$$
\partial_t\tau _{t,0}(A)=\delta (t)\big(\tau _{t,0}(A)\big)
$$ 
with initial condition $\tau_{0,0}(A)=A$, where
\begin{equation}
\delta (t):=\delta _{\mathrm{s}}+\delta _{\mathrm{c}}(t)+\delta _{\mathrm{r}%
}+\lambda \delta _{\mathrm{s},\mathrm{r}}  \label{derivation full nija}
\end{equation}
is a time-dependent, symmetric derivation of the interacting and control-driven total system. Here, $\lambda \in \mathbb{R}$ is a (small) coupling constant and the interaction with the reservoir is
\begin{equation}
\delta _{\mathrm{s},\mathrm{r}}:=i\,\left[ Q\otimes \Phi (f),\;\cdot \right] ,  \label{Macarena derivation}
\end{equation}%
where $Q$ is a self-adjoint operator on ${\mathbb{C}}^{d}$ and $\Phi(f)=\frac{1}{\sqrt{2}}[a^{\ast }(f)+a(f)]$ is the field
operator (with $f\in L^{2}({\mathbb{R}}^{3})$ a ``form factor''). Note that $\delta_{\mathrm{s,r}}$ is a bounded operator because the fields obey the Fermi statistics. 

We make the following assumptions.
\begin{itemize}
\item[(A1)] We have $T\Vert H_{\mathrm{s}}\Vert <\pi/2 $.

\item[(A2)] 
For a form factor $f\in L^2({\mathbb R}^3)$ define $g_f\in L^2({\mathbb R}\times S^2)$ by
\begin{equation}
g_f(p,\vartheta ):=|p|\left( 1+\mathrm{e}^{-\beta p}\right)
^{-1/2}\left\{ 
\begin{array}{lcr}
f(p\vartheta ) & , & p\geq 0\ , \\ 
\overline{f(-p\vartheta )} & , & p<0.
\end{array}
\right. 
\label{fg1}
\end{equation}
We assume that $g_f$ and $\widetilde g_f(p,\vartheta):=i\overline{g_f(-p,\vartheta)}$ have analytic $L^2(S^2)$-valued continuations, in the variable $p$, to the
strip $\mathbb{R+}i(-r_{\mathrm{max}},r_{\mathrm{max}})$ for some $r_{%
\mathrm{max}}>8\Vert H_{\mathrm{s}}\Vert $. Furthermore,
\begin{equation}
\sup_{|\theta |<r_{\mathrm{max}}}\int_{\mathbb{R}\times S^{2}}(1+p^{2})\big(%
\vert g_f(p+i\theta ,\vartheta )\vert
^{2}+\vert  \mathrm{e}^{-\frac{\beta}{2}(p+i\theta)}\widetilde g_f(p+i\theta ,\vartheta )\vert^{2}\big) \d p\d \vartheta <\infty .  
\label{m34}
\end{equation}
\end{itemize}

The method of (translation) analytic spectral deformation goes back to \cite{JP1, AH} and requires the technical condition (A2) on the form factor \color{black}- see Remark~\ref{rem-1.4,1}\color{black}. Condition (A1) guarantees that
resonances in this spectral deformation method do not overlap (see the discussion before \fer{61.5}). We also assume the following \textit{Dynamical Decoupling Condition},
\begin{equation}
\int_{t}^{T+t}V^*_{\mathrm{c}}(s)QV_{\mathrm{c}}(s)\mathrm{d}s=0, \qquad \mbox{for all $t\in{\mathbb R}$.}
\label{dyn.cond}
\end{equation}
This condition has already appeared in  \cite{Facchi,Vio}. It forces the system-reservoir interaction to allow for energy exchanges, for in the opposite case, the integral is simply $T Q$, which vanishes only in trivial cases. We refer to Remarks~\ref{rem-1.2}, \ref{rem-1.3} for further explanations about the dynamical decoupling condition.

We consider initial states
\begin{equation}
\omega _{0}=\omega _{\mathcal{S}}\otimes \omega _{\mathcal{R}},
\label{initialstate}
\end{equation}%
where $\omega _{\mathcal{S}}$ is any state on $\mathcal{B}({\mathbb{C}}^{d})$
and $\omega _{\mathcal{R}}$ is the equilibrium state of the reservoir, determined by (\ref{gibbsreservoir}). Here is our main result.

\begin{satz}[Effective system dynamics]
\label{mainthm} Suppose the $T$-periodic control term $H_{\mathrm{c}}$
satisfies the dynamical decoupling condition (\ref{dyn.cond}). Then there
are constants $0<c,C<\infty $, independent of $t,\lambda ,T,H_{\mathrm{c}}$,
and $\omega _{\mathcal{S}}$, such that 
\begin{eqnarray}
\lefteqn{\left\vert \omega _{0}\left( \tau _{t,0}(A\otimes \mathbf{1}_{%
\mathcal{R}})\right) -\omega _{\mathcal{S}}\left( \tau_{t,0}^{\rm s,c}(A)\right) \right\vert }  \notag \\
&&\qquad \qquad \qquad \leq C\Vert A\Vert \Big( |\lambda |+(D|\lambda
|+1)T+1-\mathrm{e}^{-ct|\lambda |T}\Big)   \label{m-3}
\end{eqnarray}%
for all $t\geq 0$, $A\in {\cal B}(\mathbb{C}^{d})$ and all initial
states $\omega _{0}=\omega _{\mathcal{S}}\otimes \omega _{{\mathcal{R}}}$.
Here, 
\begin{equation}
D:=T\max_{t\in \lbrack 0,T]}\Vert H_{\mathrm{c}}(t)\Vert .  \label{mD}
\end{equation}
\end{satz}

We point out that $H_{\mathrm c}(t)$ is of size $T^{-1}$, for $T$ small (see Remarks~\ref{rem-1.0,1}, \ref{rem-1.2}, and \ref{rem-1.3} below), hence the definition of $D$, \fer{mD}. The bound \fer{m-3} shows that up to times $t< (c|\lambda |T)^{-1}$, the
reduced dynamics of the system is approximated by the automorphism group $\tau^{\rm s,c}_{t,0}(A) = V_{\mathrm{c%
}}(t)\mathrm{e}^{itH_{\mathrm{s}}}A\mathrm{e}^{-itH_{\mathrm{s}}}V_{\mathrm{c%
}}(t)^{\ast }$ on $\mathcal{B}({\mathbb{C}}^{d})$, which is the dynamics of the periodically driven system not subjected to the interaction with the reservoir, see \fer{scdyn}. 

Since $V_{\mathrm{c}%
}(t)$ commutes with $H_{\mathrm{s}}$, the approximated dynamics leaves
eigenspaces of $H_{\mathrm{s}}$ invariant. In particular, if $E$ is a simple
eigenvalue of $H_{\mathrm{s}}$ with eigenvector $\varphi $, then $H_{\mathrm{%
c}}(t)\varphi =\Theta(t)\varphi $ for some real $\Theta(t)$, and hence
(see (\ref{vc})) $V_{\mathrm{c}}(t)\varphi =\mathrm{e}^{i\int_{0}^{t}\Theta(s)\mathrm{d}s}\varphi $. We thus obtain the following result for the dynamics of the reduced density matrix $\rho_{\mathrm{s}}(t)$, which is defined by ${\rm Tr}(\rho_{\mathrm{s}}(t) A) =\omega_0(\tau_{t,0}(A\otimes\mathbf{1}_{\mathcal{R}}))$.
\begin{koro}[Suppression of decoherence]
\label{decocoro} Suppose the
eigenvalues $\{E_k\}_{k=1}^d$ of $H_{\mathrm{s}}$ are simple, with orthonormal 
eigenvectors $\{\varphi _{k}\}_{k=1}^d$. Then 
\begin{eqnarray*}
\left\langle \varphi _{m},\rho _{\mathrm{s}}(t)\varphi _{n}\right\rangle &=&%
\mathrm{e}^{-it(E_m-E_n)+i[\Theta _{m}(t)-\Theta _{n}(t)]}\left\langle
\varphi _{m},\rho _{\mathrm{s}}(0)\varphi _{n}\right\rangle \\
&& +\mathcal{O}(|\lambda|+T+1-\mathrm{e}^{-ct|\lambda|T}),
\end{eqnarray*}
where the $\Theta_k(t)$ are real-valued and depend only on $H_{\mathrm c}(t)$.
\end{koro}

\begin{bemerkung} \label{rem-1.0,1} 
Corollary~\ref{decocoro} shows that, for small $|\lambda|$ and $T$,
decoherence is suppressed for times $t<(c|\lambda|T)^{-1}$: the
off-diagonal density matrix elements do not decay on this time-scale
(modulo a small error of size $|\lambda|+T$). On the same time-scale,
populations (diagonal elements) are constant, modulo small errors.

\color{black} The time scale $|\lambda|^{-1} T^{-1}$ is is to be
compared to the usual time scale $\lambda^{-2}$ for decoherence and
thermalization deriving in the limit of weak coupling.  We hence
conclude that the control term induces a noticable reduction of
decoherence provided that
\begin{equation} \label{eq-rem-1.0,1}
T \ \ll \ |\lambda|.
\end{equation}
\end{bemerkung}
\color{black} 
\begin{bemerkung} \label{rem-1.1} 
If the spectral subspaces of $H_\S$ are not one-dimensional, then the statement of the corollary is slightly modified. The reduced density matrix evolves ``clusterwise'': matrix elements belonging to a given cluster, i.e., those associated to a given spectral subspace, evolve jointly. Different clusters evolve independently \cite{dec1,dec2,dec3}.
\end{bemerkung}
\begin{bemerkung} \label{rem-1.2} 
The dynamical decoupling condition \eqref{dyn.cond} implies
$H_{\mathrm{c}}\gtrsim T^{-1}$. Indeed, if $Q\neq 0 $, then we obtain
from (\ref{dyn.cond}) $\max_{0\leq t\leq T} \left\Vert
V_{\mathrm{c}}(t)-\mathbf{1}\right\Vert \geq -1+\sqrt{2}$, uniformly
in $T$ and $Q$ (write $Q$ as the sum of four terms using $1=V+(1-V)$
and integrate). Thus (\ref{vc}) gives $\exp\{T\max_{0\leq t\leq T}
\|H_{\mathrm{c}}(t)\|\}-1\geq -1+\sqrt{2}$. This implies
\begin{equation} \label{eq-rem-1.2}
\max_{0 \leq t \leq T} \|H_{\mathrm{c}}(t) \|
\ \geq \ 
\frac{\ln(2)}{2T} .
\end{equation}
\end{bemerkung}
\begin{bemerkung} \label{rem-1.3} 
\color{black} Let $T,T^{\prime }>0$. It is easy to see that the
$T$-periodic $H_{\mathrm{c}}(t)$ satisfies (\ref{dyn.cond}) with $T$
if and only if the $T^{\prime }$-periodic $H_{\mathrm{c}}^{\prime
}(t)=\frac{T}{T^{\prime }}H_{\mathrm{c}}(tT/T^{\prime })$ satisfies
(\ref{dyn.cond}) with $T^{\prime }$.  We also have $T\max_{1\leq t\leq
  T}\Vert H_{\mathrm{c}}(t)\Vert =T^{\prime}\max_{1\leq t\leq
  T^{\prime }}\Vert H_{\mathrm{c}}^{\prime }(t)\Vert$. Therefore, by
comparing with $T^{\prime }=1$, we see that $T\max_{1\leq t\leq
  T}\Vert H_{\mathrm{c}}(t)\Vert =\mathrm{const}$, independent of
$T$. So the estimate $\max_{1\leq t\leq T}\Vert H_{\mathrm{c}}(t)\Vert
=\mathcal{O} (T^{-1})$ is sharp.
\end{bemerkung}
\begin{bemerkung} \label{rem-1.4} 
Let $Q(t):=V^*_{\rm c}(t) QV_{\rm c}(t)$, $t\in\mathbb R$. In our analysis, we require the two conditions
\begin{equation}
Q(T)=Q \mbox{\quad and\quad}  \widehat Q(0):=\frac1T\int_0^T Q(s)\d s=0.
\label{twocond}
\end{equation}
The condition $Q(T)=Q$ is equivalent to $Q(t+T)=Q(t)$ for all $t\in\mathbb R$. Indeed, due to the periodicity of $H_{\rm c}(t)$ (see \fer{perc}), both functions $t\mapsto Q(t+T)$ and $t\mapsto Q(t)$ satisfy the same differential equation $\frac{\d}{\d t} X(t)=-i[H_{\rm c}(t),X(t)]$ and thus the initial condition determines the solution uniquely. We show now that \fer{dyn.cond} and \fer{twocond} are equivalent. Assume that \fer{twocond} holds. Let $f(t)=\int_t^{t+T}Q(s)\d s$. Then $f'(t)=Q(t+T)-Q(t)=0$ and therefore, $f(t)=f(0)=0$, which implies condition \fer{dyn.cond}. Therefore \fer{twocond} implies \fer{dyn.cond}. Conversely, assume that \fer{dyn.cond} holds. By taking the derivative w.r.t. $t$ at $t=0$, we obtain the first equality in \fer{twocond}. By setting $t=0$ in \fer{dyn.cond} we obtain the second equality in \fer{twocond}. Therefore \fer{dyn.cond} implies \fer{twocond}. This shows that the dynamical decoupling condition \fer{dyn.cond} is \textit{equivalent} to the two conditions \fer{twocond}. We need $Q(T)=Q$ in order to have periodicity of the dynamics in the interaction picture, see \fer{tauIperiodiclemma}. The condition of a vanishing zero Fourier mode, $\widehat Q(0)=0$, is used in the proof of Theorem \ref{thmm2}, see \fer{m31}.
\end{bemerkung}
\begin{bemerkung} \label{rem-1.4,1} 
In many applications, the interaction is isotropic, and the form
factor $f\in L^2({\mathbb R}^3)$ is a radial function, $f(p\vartheta)
\equiv f(p)$. Then (A2) implies locally at $p=0$ that ${\mathbb R}^+
\ni p \mapsto p f(p)$ is an even, real, regular function (which can be
expanded in a power series about $p=0$).
\end{bemerkung}
\begin{bemerkung} \label{rem-1.5} 
\color{black}
We are deriving our results for fermion fields because these are
mathematically easier to deal with than boson fields, for the
interaction is a bounded operator in this case. In contrast, the
unboundedness of the interaction operator in case of boson fields
brings along several additional difficulties: The existence of the
unitary time-evolution operator is not trivial anymore, especially,
not for the explicitly time-dependent perturbation we are considering
here. Analyticity of the family of Liouvilleans under consideration
w.r.t.\ complex translation of the field momenta is not easy to see,
either.

In anticipation of a very similar analysis for bosons, the Hamiltonian
we study has the form of the spin-boson Hamiltonian - the model for
the main application we have in mind -, but the fields fulfill the
canonical anticommutation relations, rather than canonical commutation
relations. With this replacement, however, the model as such is
physically not meaningful because it mixes the even and the odd sectors
of the fermion algebra. We note, however, that it is easy to extend
our analysis from $\delta _{\mathrm{s},\mathrm{r}} = i\,[ Q\otimes
  \Phi (f), \; \cdot ]$ in \eqref{Macarena derivation} to $\delta
_{\mathrm{s},\mathrm{r},\cP} := i\,[ Q\otimes \cP(\Phi), \; \cdot ]$,
where $\cP(\Phi)$ is an arbitrary real polynomial in the field
$\Phi$. We consider a coupling linear in $\Phi$ to keep technical
aspects as simple as possible. See also
\cite[1.3. Remarks.]{JaksicPillet}.
\end{bemerkung}
\begin{bemerkung} \label{rem-1.6} 
\color{black} On a technical level, Eq.~\eqref{force-commute} is a key assumption
for the derivation of our results, as it allows to move the explicit
time-dependence of the forcing, which is big in magnitude, into an
explicit time-dependence of the interaction, which is small in
magnitude. Physically, if the small system is an atom held in a
magnetic trap and the forcing is a magnetic force, this means that the
directions of these two magnetic fields are parallel. 
It would be
desirable, however,
 to also treat the case in which the forcing is
perpendicular to the magnetic trap field. We do not study that case in
the present paper.
\color{black}
Indeed, note that a forcing parallel to the trap field corresponds to a control term $H_{\mathrm{c}}(t)$
commuting with the atom Hamiltonian $H_{\mathrm{s}}$. In this situation the populations (diagonal terms of the density matrix) of the
initial state are not affected by the forcing. By contrast, if the forcing  is perpendicular to the trap field, then $H_{\mathrm{c}}(t)$
does not commute with $H_{\mathrm{s}}$ and even a very small forcing can cause drastic changes of populations
of the atom, by a so-called ``pumping" mechanism, at large enough times (as compared to $\|H_{\mathrm{c}}(t)\|^{-1}$). 
Hence, in this case, the problem of conservation of quantum information is more subtle.    
\end{bemerkung}
\color{black}


\noindent
\textbf{Organization of the paper.} In Section \ref{outline proof} we
outline the main steps of the proof of Theorem \ref{mainthm}. Technical
details are presented in Section \ref{section proof}. In Section \ref{sect
example} we apply our results to the spin-fermion model. We obtain explicit expressions for the times until which coherence is preserved (the `controlled decoherence times'). They coincide with those found by formal computations in \cite{Facchi}.

\subsection{Outline of proof of Theorem \protect\ref{mainthm}\label{outline
proof}}

The purpose of this section is to explain the main four steps in the proof of Theorem \ref{mainthm}. The details are given in Section \ref{sectproof}.

\subsubsection{Interaction picture}

By passing to the interaction picture, we transfer the dependence on the
period $T$, which is considered a small parameter, into the interaction
term. To this end, let $\mathfrak{H}_{\mathrm{s}}$ be the Hilbert space $\mathcal{B}(\mathbb{C}^{d})$ endowed with the Hilbert--Schmidt scalar
product $\left\langle A,B\right\rangle _{\mathrm{s}}:=\mathrm{Tr}(A^{\ast
}B) $, for $A,B\in \mathcal{B}(\mathbb{C}^{d})$. Define the unitaries $\widetilde V_{%
\mathrm{c}}(t)$ on $\mathfrak{H}_{\mathrm{s}}$ by 
\begin{equation}
\widetilde V_{\mathrm{c}}(t)A:=V_{\mathrm{c}}(t)^{\ast }AV_{\mathrm{c}}(t),\text{ \ \ }%
A\in \mathfrak{H}_{\mathrm{s}},\ t\in {\mathbb{R}},  \label{m28}
\end{equation}
where $V_{\mathrm c}(t)$ is given in \fer{vc}. 
We extend $\widetilde V_{\mathrm{c}}(t)$ to all of ${\mathcal{V}}=\mathcal{B}(\mathbb{C%
}^{d})\otimes {\mathcal{V}}_{\mathcal{R}}$ by trivial action on ${\mathcal{V}%
}_{\mathcal{R}}$ and set, for $s\leq t\in {\mathbb{R}}$, 
\begin{equation}
\tau _{t,s}^{I}:=\widetilde V_{\mathrm{c}}(t)\circ\tau _{t,s}\circ\widetilde V_{\mathrm{c}}(s)^{\ast }.
\label{m-1}
\end{equation}
We have
$$
\partial_t \tau^I_{t,s}(A) =\delta^I_t\left(\tau_{t,s}^I(A)\right),
$$
where the time-dependent symmetric derivation is
\[
\delta _{t}^{I}:=\delta _{\mathrm{s}}+\delta _{\mathrm{r}}+\lambda \delta _{%
\mathrm{s},\mathrm{r}}^{I}(t),\quad t\in \mathbb{R},
\]%
where%
\begin{eqnarray}
\delta _{\mathrm{s},\mathrm{r}}^{I}(t)&:=&i\,\left[ Q(t)\otimes \Phi
(f),\,\cdot \, \right] ,  \notag \\
Q(t)&:=&V_{\mathrm{c}}(t)^{\ast }QV_{\mathrm{c}}(t).  \label{qoft}
\end{eqnarray}%
The dynamical decoupling condition (\ref{dyn.cond}) implies that $Q(t+T)=Q(t)$ (see \fer{twocond}) and consequently, $\tau^I$ is $T$-periodic, i.e.,
\begin{equation}
\tau _{t+T,s+T}^{I}=\tau _{t,s}^{I},\mbox{\quad $t\geq s$}.
\label{tauIperiodiclemma}
\end{equation}

\subsubsection{Howland-Yajima Hilbert space}

We represent the total system on its Gelfand-Naimark-Segal (GNS) Hilbert
space $\mathfrak{H}$ with inner product $\left\langle {\cdot },{\cdot }%
\right\rangle_{\mathfrak H}$, 
\begin{equation}
\omega _{0}(\tau _{t,0}^{I}(A))=\left\langle {\Omega _{0}},{U_{t,0}\pi
(A)\Omega _{0}}\right\rangle_{\mathfrak H}.
\label{pr0}
\end{equation}
Here, $\Omega _{0}$ is the vector in $\mathfrak{H}$ representing the initial state $\omega _{0}$, see (\ref%
{initialstate}), $\pi $ is the representation
map (sending observables to bounded operators on $\mathfrak{H}$), and $%
U_{t,0}$ is the implemented periodic dynamics, satisfying $%
U_{t+T,s+T}=U_{t,s}$ (see Theorem \ref{corolario legal copy(1)}). Next, we
pass to the Howland-Yajima Hilbert space 
$$
\mathfrak{H}_{\mathrm{per}}=L_{\mathrm{per}}^{2}({\mathbb R}/T{\mathbb Z},{\mathfrak{H}})
$$ 
of $T$-periodic, $\mathfrak{H}$-valued functions, with inner product 
$$
\left\langle {\psi },{\chi }\right\rangle _{\mathrm{per}}=\frac{1}{T}\int_{0}^{T}\left\langle {\psi (x)},{\chi (x)}\right\rangle_{\mathfrak H}\d x
$$
\cite{How,Yaji1,Yaji2}. We denote ${\mathbb T}_T:={\mathbb R}/T{\mathbb Z}$. For all $t\geq s$, we define $U^{\rm per}_{t,s}$ acting on ${\mathfrak H}_{\rm per}$ by
$$
(U^{\rm per}_{t,s}\psi)(x) = U_{t,s}(\psi(x)),\qquad \psi\in{\mathfrak H}_{\rm per},\ x\in{\mathbb T}_T.
$$
Moreover, by 
$$
Z:{\mathfrak H}\rightarrow{\mathfrak H}_{\rm per}, \qquad (Z\psi)(x):=\psi,
$$
we map vectors in $\mathfrak H$ to constant $\mathfrak H$-valued functions on ${\mathbb T}_T$. Then \fer{pr0} reads
$$
\omega_0(\tau _{t,0}^{I}(A))=\left\langle Z\Omega_{0},U^{\rm per}_{t,0} Z(\pi
(A)\Omega_{0})\right\rangle_{\rm per}.
$$
The Howland, or Floquet generator $\mathcal{G}$ is defined on ${\mathfrak H}_{\rm per}$ by 
\begin{equation}
(\mathrm{e}^{i\alpha \mathcal{G}}\psi )(t)=U^{\rm per}_{t,t-\alpha }\psi (t-\alpha ),
\label{howgen}
\end{equation}
the r.h.s. being a strongly continuous semigroup in $\alpha \geq 0$. In the
remaining part of this section, we denote by $A$ a system observable, acting
on $\mathfrak{H}_{\mathrm{s}}$ (instead of writing $A\otimes \mathbf{1}_{%
\mathcal{R}}$). We have (Lemma \ref{lemmalongtime}) 
\begin{equation}
\left\vert \left\langle {\Omega _{0}},{U_{\alpha ,0}\pi (A)\Omega _{0}}
\right\rangle_{\mathfrak H} -\left\langle {Z\Omega _{0}},{\mathrm{e}^{i\alpha \mathcal{G}}Z(\pi (A)\Omega _{0})}\right\rangle _{\mathrm{per}}\right\vert =\mathcal{O}(T).
\label{pr1}
\end{equation}
This estimate can be understood as follows. Since 
$$
\langle\Omega_0, U_{\alpha,0}\pi (A)\Omega _{0}\rangle_{\mathfrak H}= \langle Z\Omega_0, U^{\rm per}_{\alpha,0}Z(\pi(A)\Omega_0)\rangle_{\rm per},
$$
the left side of \fer{pr1} is given by $\frac{1}{T}\int_{0}^{T}\left\langle{\Omega _{0}},{(U_{\alpha ,0}-U_{t,t-\alpha })\pi (A)\Omega _{0}}%
\right\rangle_{\mathfrak H} \d t$, and the difference $(U_{\alpha
,0}-U_{t,t-\alpha })\pi (A)\Omega _{0}$ is small, for small $T$. That is, given $\alpha $ and $t$, we have $t+nT=\alpha +\epsilon $ for some integer $%
n $ and some $0\leq \epsilon <T$, and so by periodicity, $U_{t,t-\alpha }\pi
(A)\Omega _{0}=U_{\alpha +\epsilon ,\epsilon }\pi (A)\Omega _{0} =
U_{\alpha ,0}\pi (A)\Omega _{0}+O(\epsilon)$.

\subsubsection{Spectral deformation, resonances, extraction of main term}

Let $\mathcal{G}_0$ denote the Howland generator in (\ref{howgen}), for $%
\lambda=0$ (no system-reservoir interaction). We show in Theorem \ref%
{section extension copy(4)} that 
\begin{equation}  \label{pr2}
\left|\left\langle {Z\Omega_0}, {(\mathrm{e}^{i\alpha \mathcal{G}_0} -\mathrm{e%
}^{i\alpha\mathcal{G}})Z(\pi(A)\Omega_0)}\right\rangle_{\mathrm{per}}\right| =%
\mathcal{O}(|\lambda|+T+\mathrm{e}^{\alpha c|\lambda|T}-1).
\end{equation}
We explain how to arrive at this bound. The Fourier transform $\frak F$ with respect to $t$,
\begin{equation}
({\frak F}\psi)(k)\equiv \widehat\psi(k) :=\frac1T\int_0^T {\rm e}^{-2\pi i kt/T}\psi(t)\d t,
\label{FT}
\end{equation}
where $k\in\mathbb Z$, maps $\mathfrak H_{\rm per}$ into ${\frak F}{\mathfrak H}_{\rm per}{\frak F}^*=  \widehat{\mathfrak{H}}_{\mathrm{per}}=l^2({\mathbb{Z}},\mathfrak{H})$. We denote the inner product of $\widehat{\mathfrak{H}}_{\mathrm{per}}$ by $\langle\cdot,\cdot\rangle_{\widehat{\rm per}}$ and set $\widehat Z={\frak F}\circ Z:{\mathfrak H}\rightarrow \widehat{\mathfrak{H}}_{\mathrm{per}}$. By unitarity of the Fourier transform and deformation analyticity
(translations), it suffices to prove the bound \fer{pr2} in Fourier space and for the
spectrally deformed generators $\widehat{\mathcal{G}}_0(\theta)$ and $%
\widehat{\mathcal{G}}(\theta)$, where 
\begin{equation}  \label{pr3}
\widehat{\mathcal{G}}_0(\theta) :={\frak F} {\mathcal G}_0(\theta){\frak F}^*= -\frac{2\pi}{T}k + L+\theta\widehat N.
\end{equation}
Here, $k$ is the operator of multiplication by the argument in $l^2({\mathbb{%
Z}},\mathfrak{H})$, $L=L_{\rm s}+L_\R$ is the self-adjoint Liouvillian on $\mathfrak{H}$ (acting
fiber-wise on $l^2({\mathbb{Z}},\mathfrak{H})$, for each $k$), associated to the free motion of the system and of the reservoir. We have $L_{\rm s}=[H_{\rm s},\cdot]$ (commutator) and $L_\R$ has a simple eigenvalue at the origin (the associated eigenvector being the KMS state), embedded in continuous spectrum covering all of $\mathbb R$. The $\widehat
N $ is a ``number operator'', acting on $\mathfrak{H}$ (fiber-wise),
commuting with $L$ and having spectrum $\{0,1,2,\ldots\}$. The spectrum of $\widehat{\mathcal{G}}_0(\theta)$ consists of eigenvalues $\frac{2\pi}{T}{\mathbb{Z}} +\mathrm{spec}([H_{\mathrm{s}},\cdot])$ on the real  axis
and continuous spectrum on the horizontal lines ${\mathbb R}+i{\mathbb N}\,{\rm Im}\theta$. For $\mathrm{Im} \,\theta>0$, the branches of continuous spectrum are pushed into the upper complex half-plane ${\mathbb C}^+$, moved away from the
eigenvalues on the real axis. Eigenvalues
of $\widehat{\mathcal{G}}_0(\theta)$ associated to different $k$ (see (\ref%
{pr3})) are separated from each other due to the condition (A1) of
non-overlapping resonances (see before Theorem \ref{mainthm}). For $|\lambda|$ small, this separation persists by perturbation theory. Denote by $%
\mathcal{P}_k(\lambda)$ the Riesz projection of $\widehat{\mathcal{G}}(\theta)$ associated to the group of real eigenvalues associated to $k\in{%
\mathbb{Z}}$, i.e., those lying in the vicinity of $\frac{2\pi}{T}k$. By the Laplace inversion formula, $\mathrm{e}^{i\alpha\widehat{%
\mathcal{G}}(\theta)}$ is expressed as a complex line integral over the
resolvent $(\mathcal{G}(\theta)-z)^{-1}$ along the horizontal path ${\mathbb R}-i$.
Deforming this contour to the horizontal line $z={\mathbb{R}}+ir/2$, where $0<r\equiv \mathrm{Im} \, \theta< r_{\mathrm{max}}$, one generates residue
contributions at the eigenvalues of $\mathcal{G}(\theta)$, which migrate (with varying $\lambda$) into the upper complex half-plane, 
\begin{eqnarray} 
\left\langle {\widehat Z\Omega_0}, {\mathrm{e}^{i\alpha\widehat{\mathcal{G}}}\widehat Z(\pi(A)\Omega_0)}\right\rangle_{\widehat{\rm{per}}} &=& \sum_{k\in{\mathbb{Z}}} \left\langle\widehat Z
\Omega_0, \mathrm{e}^{i\alpha \widehat{\mathcal{G}}(ir)} \mathcal{P}_k(\lambda)\widehat Z(\pi(A)\Omega_0)\right\rangle_{\widehat{\rm{per}}}\nonumber\\
&& +\mathcal{O}(\lambda^2\mathrm{e}^{-\alpha
r/2}).
\label{pr4}
\end{eqnarray}
The series on the right side of \fer{pr4} converges since 
\begin{equation}  \label{pr5}
\|\mathcal{P}_k(\lambda)\widehat Z(\pi(A)\Omega_0)\| \leq C\left(\frac{T}{k}%
\right)^2\max_{0\leq t\leq T}\|H_{\mathrm{c}}(t)\| =k^{-2} {\mathcal O}(T).
\end{equation}
To understand the bound \fer{pr5}, we note that $\widehat Z(\pi(A)\Omega_0)$ belongs to the domain
of $\widehat{\mathcal{G}}(\theta)^2$, so 
$$
\|\widehat{\mathcal{G}}(\theta)^2\widehat Z(\pi(A)\Omega_0)\|^2_{\widehat{\rm per}} \approx \sum_{k\in{%
\mathbb{Z}}} \left(\frac{2\pi k}{T}\right)^4 \|P_k(\lambda)\widehat Z(\pi(A)\Omega_0)\|^2_{\widehat{\rm per}}
$$ 
converges, which implies the decay $\|P_k(\lambda)\widehat Z(\pi(A)\Omega_0)\|_{\widehat{\rm per}}\sim (T/k)^2$. The bound (\ref{pr5}) implies that, for small $T$, the main contribution to the
sum in (\ref{pr4}) comes from $k=0$, 
\begin{equation}  \label{pr6}
\left\langle \widehat Z\Omega_0, \mathrm{e}^{i\alpha\widehat{\mathcal{G}}}\widehat Z(\pi(A)\Omega_0)\right\rangle_{\widehat{\rm per}} = \left\langle \widehat Z\Omega_0, \mathrm{e}^{i\alpha \widehat{\mathcal{G}}(ir)}\mathcal{P}_0(\lambda)\widehat Z(\pi(A)\Omega_0)\right\rangle_{\widehat{\rm per}} +\mathcal{O}(\lambda^2+T).
\end{equation}
The final step in showing (\ref{pr2}) is the estimate
\begin{equation}  \label{pr7}
\mathrm{e}^{i\alpha\widehat{\mathcal{G}}_0(ir)}\mathcal{P}_0(0)- \mathrm{e}%
^{i\alpha\widehat{\mathcal{G}}(ir)}\mathcal{P}_0(\lambda)=\mathcal{O}%
(|\lambda| +\mathrm{e}^{c\alpha|\lambda|T}-1).
\end{equation}
We show \fer{pr7} in Theorem \ref{thmm2} by comparing, with the help of Kato's method of similar projections, the two dynamics generated by $\widehat{\mathcal{G}}(ir)\mathcal{P}_0(\lambda)$ and $\widehat{\mathcal{G}}_0(ir)\mathcal{P}_0(0)$, respectively. The proof of this theorem uses the vanishing of the zero Fourier mode, $\widehat Q(0)=0$, see also \fer{twocond}. 

\subsubsection{Proof of Theorem \protect\ref{mainthm}}

Putting together (\ref{pr0}), (\ref{pr1}) and (\ref{pr2}) we obtain 
\begin{equation}  \label{pr10}
\omega_0(\tau^I_{t,0}(A)) = \left\langle \widehat Z\Omega_0, \mathrm{e}^{i t {\mathcal G}_0}\widehat Z(\pi(A)\Omega_0)\right\rangle_{\mathrm{per}} 
+\mathcal{O}\big(|\lambda|+T+\mathrm{e}^{c t |\lambda| T}-1\big).
\end{equation}
We have $\tau^I_{t,0}(A)=V_{\mathrm{c}}(t)^*\tau_{t,0}(A) V_{\mathrm{c}}(t)$
and $\mathrm{e}^{it\mathcal{G}_0}\pi(A)\Omega_0 = \pi( \mathrm{e}^{itH_{%
\mathrm{s}}}A \mathrm{e}^{-itH_{\mathrm{s}}})\Omega_0$. Finally, since the
l.h.s. and the first term on the r.h.s. of (\ref{pr10}) are bounded
uniformly in $t$, we may replace $\mathrm{e}^{c t |\lambda| T}-1$ in the
remainder term by $\min\{1,\mathrm{e}^{c t |\lambda| T}-1\}$. The latter
minimum is bounded above by $2(1-\mathrm{e}^{-c t |\lambda|T})$. The bound (\ref{m-3}) follows.

\section{Spin-fermion model\label{sect example}}

{}For the concrete model of a two-level system (`qubit') for our system, we are able to obtain more detailed information on the dynamics. We present the analysis in this section, and we recover a bound on the decoherence time that has been derived in the physics literature by heuristic means before in \cite{Facchi}.

The Hilbert space of the spin is ${\mathbb{C}}^{2}$, on which the spin Hamiltonian $H_{\mathrm{s}}$, the forcing term $H_{\mathrm{c}}(t)$, and the interaction operator $Q$ act as
\begin{equation*}
H_{\mathrm{s}}=\left[ 
\begin{array}{ll}
1 & 0 \\ 
0 & -1%
\end{array}%
\right] ,\quad H_{\mathrm{c}}(t)=\frac{\mu}{T}\varkappa (t/T)\left[ 
\begin{array}{ll}
1 & 0 \\ 
0 & -1%
\end{array}%
\right] ,\quad Q=\left[ 
\begin{array}{ll}
0 & 1 \\ 
1 & 0%
\end{array}%
\right],
\end{equation*}
respectively. Here, $\mu$ is a real coupling constant and  $\varkappa (\cdot)$ is a smooth, real function of period one. 

\medskip
{\bf Example.\ } The choice $\varkappa (t)=\cos (2\pi t)$ gives 
\begin{equation}
Q(t)=\left[ 
\begin{array}{cc}
0 & \exp (-\frac{i\mu }{\pi }\sin (\frac{2\pi }{T}t)) \\ 
\exp (\frac{i\mu }{\pi }\sin (\frac{2\pi }{T}t)) & 0%
\end{array}%
\right] \otimes \Phi (\mathrm{f}),
\end{equation}
see \fer{qoft}, and the dynamical decoupling condition (\ref{dyn.cond}) becomes
\begin{equation}
\int_{-\pi }^{\pi }\cos \left[ \frac{\mu }{\pi }\sin (s)\right] \mathrm{d}%
s=0.  \label{ddc.example}
\end{equation}
A simple stationary phase argument shows that, as $\mu$ varies throughout $\mathbb R$, the set of real numbers $\int_{-\pi }^{\pi }\cos \left[ \frac{\mu }{\pi }\sin (s)\right] \mathrm{d} s$ contains negative and positive values. In particular, by continuity, \fer{ddc.example} holds for some $\mu$. (Condition \fer{ddc.example} is actually related to the existence of zeros of certain Bessel functions.)


\medskip
We will assume from now on that $\varkappa$ is chosen so that the dynamical decoupling condition is satisfied, but we do not have to take it as in the example. We now show how one can obtain finer information about the dynamics than the one given in Theorem \ref{mainthm}, by choosing a ``corrected'' uncoupled dynamics. The decoherence time estimate comes from  an estimate on the difference of the total dynamics to a coherence preserving one, $\mathrm{e}^{i\alpha \widehat{\mathcal{G}}_{0}(ir)}\mathcal{P}_{0}(0)-\mathrm{e}^{i\alpha \widehat{\mathcal{G}}(ir)}\mathcal{P}_{0}(\lambda )$, see \fer{pr7}. The idea now is that by modifying $\widehat{\mathcal{G}}_{0}(ir)$ to $\widehat{\mathcal{G}}_{0}(ir)+\Delta$, where $\Delta=[S,\cdot]$ with a self-adjoint operator $S$ \textit{which commutes with $H_{\mathrm{s}}$}, one can achieve two things. Firstly, coherences are still preserved, as $S$ commutes with $H_{\mathrm{s}}$ -- in Corollary \ref{decocoro}, $S$ simply changes the phases. Secondly, the difference between the generators $\widehat{\mathcal{G}}_{0}(ir)+\Delta$ and $\widehat{\mathcal{G}}(ir)$ becomes smaller for an appropriate $\Delta$, which gives a sharper estimate on the difference of the dynamics they generate. We now implement this idea.

Recall that $\widehat{{\mathcal{G}}}=\widehat{{\mathcal{G}}}(ir)$ is the
spectrally deformed Howland operator and $\mathcal{P}_{0}(\lambda )$ its
Riesz projection associated to the group of discrete eigenvalues emerging from the eigenvalue $0$ of $\widehat{\cal G}_0=\widehat{\cal G}_0(ir)$, as $\lambda\neq 0$.
The map $\lambda \mapsto \widehat{{\mathcal{G}}}\mathcal{P}_{0}(\lambda )$ is analytic at $\lambda = 0$. We write 
\begin{equation}
\widehat{{\mathcal{G}}}\mathcal{P}_{0}(\lambda )-\widehat{{\mathcal{G}}}_{0} \mathcal{P}_{0}(0)=\sum_{m\geq 1} \lambda ^{m} A_m.
\label{mmm0}
\end{equation}
It is not hard to see that $A_m=0$ for all $m$ odd, and that $\|A_m\|\leq C^m$ for a constant $C$ independent of $0<T\leq 1$. The second order term is the \textit{level shift operator}
\[
\lambda^2 A_2 =-\lambda^2  \lim_{\varepsilon \searrow 0}  \sum_{e \in \sigma (\delta _{\mathrm{s}%
})}\ \mathcal{P}_{0,e}(0)\ \widehat{V}_{\mathrm{per}}(ir)\ (\widehat{{%
\mathcal{G}}}_{0}-e+i\varepsilon )^{-1}\ \widehat{V}_{\mathrm{per}}(ir)\ 
\mathcal{P}_{0,e }(0)
\]
(see also e.g. \cite{dec1,dec2,dec3,DerezinskiFruboes2006}). Here, we have defined, for $e\in \sigma (\delta _{\mathrm{s}})=\{-2,0,2\}$, the orthogonal projections $\mathcal{P}_{0,e }(0)$ with range 
$$
\mathrm{Ran}\,\mathcal{P}_{0,e }(0)=\{ \Psi \in \mathrm{Ran}\, \mathcal{P}_{0}(0)\text{ }:\text{ }\widehat{{\mathcal{G}}}_{0}\Psi
=e \Psi \}.
$$
We have also defined the interaction operator $\widehat{V}_{\mathrm{per}}(ir)=\widehat{{\mathcal{G}}}(ir)-\widehat{{\mathcal{G}}}_{0}(ir)$  (c.f. sections \ref{Howland} and \ref{sectionspectral deformation}). Let $A\in{\cal B}({\mathbb C}^d)$ and define $\underrightarrow{A}$ and $\underleftarrow{A}$ as the left and right multiplication on operators $B \in {\cal B}({\mathbb C}^d)$,
\begin{equation}
B\mapsto \underrightarrow{A}B:=AB\quad \text{and}\quad B\mapsto 
\underleftarrow{A}B:=BA.
\label{left multi}
\end{equation}
\color{black} 
Note that, as ${\cal B}({\mathbb C}^d)$ is isomorphic to ${\mathbb
  C}^d \otimes {\mathbb C}^d$, we may naturally identify
$\underrightarrow{A}$ and $\underleftarrow{A}$ with $A \otimes \one$
and $\one \otimes A \in {\cal B}({\mathbb C}^d \otimes {\mathbb C}^d)$,
respectively.

A straightforward, but lengthy and tedious, calculation leads us to
\color{black}
\begin{eqnarray} \label{eq-Lindblad}
\lambda ^{2}A_2 & = &
-i\frac{\lambda ^{2}}{2}\sum\limits_{a =\pm 1}
\sum\limits_{k \in \mathbb{Z} \backslash \{0\}}
\left[ \pi G_f(k/T+2a)( 2\underrightarrow{{Q}_{k,a }^{\ast }}
\underleftarrow{{Q}_{k,a }}-
\underrightarrow{{Q}_{k,a }^{\ast}}\underrightarrow{%
{Q}_{k,a }}-\underleftarrow{{Q}_{k,a }}\underleftarrow{%
\text{ }{Q}_{k,a }^{\ast }})\right.  
\notag \\
& & \left. +i \mathrm{PV}
\Big[\frac{1}{p-(k/T+2a)}\Big](G_f)
(\underleftarrow{%
{Q}_{k,a }}\underleftarrow{\text{ }{Q}_{k,a }^{\ast }}-%
\underrightarrow{{Q}_{k,a }^{\ast }}\underrightarrow{{Q}%
_{k,a }}) \right],
\label{mmm1}
\end{eqnarray}
where 
$$
{Q}_{k,a}:=\int_{0}^{1}\mathrm{e}^{-2\pi i   kt} \ 
\mathrm{e}^{-iH_{\mathrm{s}}\int_{0}^{t}\varkappa (s)\mathrm{d}s} q_{a
} \mathrm{e}^{iH_{\mathrm{s}}\int_{0}^{t}\varkappa (s)\mathrm{d}s}
\mathrm{d}t,
$$
with
\begin{equation*}
q_{-1}:=\left[
\begin{array}{ll}
0 & 1 \\ 
0 & 0%
\end{array}%
\right] ,\text{ \ \ }q_{1}:=\left[ 
\begin{array}{ll}
0 & 0 \\ 
1 & 0%
\end{array}
\right] .
\end{equation*}
Here, the function $G_f:\mathbb{R}\rightarrow \mathbb{R}_{0}^{+}$
is defined by
$$
G_f(p):=
\int_{S^{2}}\frac{p^{2}}{1+\mathrm{e}^{-\beta p}}%
\left\vert g_f(p,\vartheta )\right\vert ^{2}\mathrm{d}\vartheta
$$ 
for all $p\in \mathbb{R}$, and 
\[
\mathrm{PV} \Big[\frac{1}{p-(k/T+2a)} \Big](G_f) :=
\lim_{\varepsilon \searrow 0} 
\int_{\mathbb{R} \backslash (-\varepsilon, \varepsilon)}
\frac{G_f(p+k/T+2a)}{p} \ \mathrm{d} p
\] 
is the principal value of the function $p\mapsto \frac{G_f(p)}{p-(k/T+2a)}$ at the singularity $p=k/T +2a$.

\color{black}
\begin{bemerkung} \label{rem-2.1} 
In \eqref{eq-Lindblad} we recognize that $\lambda^2 A_2$ assumes the
standard form of a Lindblad generator, the first three terms in the
sum giving the dissipative part and the last two corresponding to the
Hamiltonian part.
\end{bemerkung}
\color{black}
\begin{bemerkung} \label{rem-2.2} 
The term $k=0$ is absent in the above sum over the $k$, due to the
dynamical decoupling condition (\ref{dyn.cond}).
\end{bemerkung}
\color{black}
The self-adjoint operators ${Q}_{k,a}^{\ast }{Q}_{k,a}$ commute with $H_{\mathrm{s}}$. We define
\begin{equation}
\Delta =\frac{\lambda ^{2}}{2}\sum\limits_{a=\pm 1}
\sum\limits_{k \in \mathbb{Z} \backslash \{0\}}  (\underleftarrow{{Q}_{k,a }}\underleftarrow{\text{ }{Q}^{\ast }_{k,a }} - \underrightarrow{{Q}_{k,a }^{\ast }}\underrightarrow{{Q}%
_{k,a }})  \mathrm{PV}\Big[\frac{1}{1- (k/T+2a)} \Big](G_f),
\label{32.1}
\end{equation}
so that, from \fer{mmm0} and \fer{mmm1}, we obtain
\begin{eqnarray}
\lefteqn{
\widehat{{\mathcal{G}}}\mathcal{P}_{0}(\lambda )-(\widehat{{\mathcal{G}}}_0+\Delta) \mathcal{P}_{0}(0)=}\label{32.2}\\
&& -i\frac{\lambda ^{2}}{2}\sum\limits_{a =\pm
1}\sum\limits_{k \in \mathbb{Z} \backslash \{0\}}
\pi G_f(k/T+2a)
 ( 2\underrightarrow{{Q}_{k,a }^{\ast }}\underleftarrow{{Q}_{k,a }}-\underrightarrow{{Q}^{\ast }_{k,a }}\underrightarrow{{Q}_{k,a }}-\underleftarrow{{Q}^{\ast }_{k,a }}\underleftarrow{%
\text{ }{Q}_{k,a }}) +\mathcal{O}(\lambda^4). \notag 
\end{eqnarray}
Therefore,
\begin{equation}
\| \widehat{{\mathcal{G}}}\mathcal{P}_{0}(\lambda )-(\widehat{{\mathcal{G}}}_0+\Delta) \mathcal{P}_{0}(0)\| \leq 2\pi\lambda^2 \xi(T) +c\lambda^4,
\label{xxy1}
\end{equation}
for some constant $c$ independent of $0<T\leq 1$, and where 
\begin{equation}
\xi(T) = \sum_{a=\pm 1}\sum_{k \in \mathbb{Z} \backslash \{0\}} 
 \|Q_{k,a}\|^2 \ |G_f(k/T+2a )|^2.
\label{xxy2}
\end{equation}
We pass from estimate \fer{xxy1} on the difference of the generators to an estimate on the difference of the propagators,
\begin{eqnarray}
\lefteqn{
\mathrm{e}^{i\alpha \widehat{{\mathcal{G}}}}\mathcal{P}_{0}(\lambda )
-\mathrm{e}^{i\alpha(\widehat{{\mathcal{G}}}_0+\Delta) }
\mathcal{P}_{0}(0) } \\
&=&
\mathrm{e}^{i\alpha \widehat{{\mathcal{G}}}\mathcal{P}_{0}(\lambda )}
\mathcal{P}_{0}(\lambda)
-\mathrm{e}^{i\alpha (\widehat{{\mathcal{G}}}_0+\Delta) \mathcal{P}_{0}(0)}
\mathcal{P}_{0}(0) \nonumber \\
&=&
(\mathrm{e}^{i\alpha \widehat{{\mathcal{G}}}\mathcal{P}_{0}(\lambda )}
-\mathrm{e}^{i\alpha (\widehat{{\mathcal{G}}}_0+\Delta)\mathcal{P}_{0}(0)})
\mathcal{P}_{0}(\lambda) +
\mathrm{e}^{i\alpha (\widehat{{\mathcal{G}}}_0+\Delta)\mathcal{P}_{0}(0)}
(\mathcal{P}_{0}(\lambda )-\mathcal{P}_{0}(0)). \nonumber
\end{eqnarray}
The norm of the second summand on the right side is bounded above by 
$C|\lambda|$, since 
$\|\mathrm{e}^{i\alpha (\widehat{{\mathcal{G}}}_0+\Delta)\mathcal{P}_{0}(0)}\|=1$ and $\mathcal{P}_{0}(\lambda )-\mathcal{P}_{0}(0)= \mathcal{O}(\lambda)$. 
By first iterating the Duhamel formula, and then using \fer{xxy1}, we obtain the estimates
$$
\| \mathrm{e}^{i\alpha \widehat{{\mathcal{G}}}\mathcal{P}_{0}(\lambda )}
-\mathrm{e}^{i\alpha (\widehat{{\mathcal{G}}}_0+\Delta)\mathcal{P}_{0}(0)} \| \leq  \mathrm{e}^{\alpha \| \widehat{{\mathcal{G}}}\mathcal{P}_{0}(\lambda )-
(\widehat{{\mathcal{G}}}_0+\Delta)\mathcal{P}_{0}(0)\|}-1 \leq 
\mathrm{e}^{2\pi\alpha \lambda^2[ \xi(T)+c\lambda^2]}-1.
$$
This gives the bound 
\begin{equation}
\left\Vert \mathrm{e}^{i\alpha (\widehat{\mathcal{G}}_{0}(ir)+\Delta )}%
\mathcal{P}_{0}(0)-\mathrm{e}^{i\alpha \widehat{\mathcal{G}}(ir)}\mathcal{P}%
_{0}(\lambda )\right\Vert   \leq  C\left( |\lambda |+
\mathrm{e}^{2\pi\alpha \lambda^2[\xi(T)+c\lambda^2]}-1\right),
\label{33}
\end{equation}
for constants $c,C$ independent of $0<T\leq 1$, $\lambda $ sufficiently small, and where $\xi(T)$ is given in \fer{xxy2}. The bound  \fer{33} shows that decoherence is suppressed for times up to 
\begin{equation}
t_{{\rm dec}}=\frac{1}{2\pi\lambda^2[\xi(T)+c\lambda^2]}.
\label{leading order dec time}
\end{equation}
{}For $\lambda^2\xi(T)+\lambda ^{4} <\!\!<| \lambda |T$, the bound \fer{leading order dec time}  on the decoherence time is {\it better} than the general one obtained in Theorem \ref{mainthm}. Also, \fer{leading order dec time} shows that to leading order in $\lambda$, the decoherence time can be very large even at moderate control frequencies. Indeed, with a suitable choice of $\varkappa$, one can achieve that $\|Q_{k,a}\| G_f(k/T+2a)$ is very small, and hence so is $\xi(T)$ (see \fer{xxy2}), driving $t_{{\rm dec}}$ in \fer{leading order dec time} to very large values. We refer to \cite{Facchi} for more discussions about this point. It has to be noted that the analysis leading to Theorem \ref{mainthm} was carried out with small values of $T$ in mind. For $T\sim 0$, $t_{{\rm dec}}$ in  \fer{leading order dec time} behaves as $1/\lambda^4$, while the bound on the decoherence time in Theorem \ref{mainthm} is $(|\lambda|T)^{-1}$, which is better (larger).

\medskip
\noindent
{\bf Bang-bang control.\ } Take $\varkappa $ to be a $1$-periodic sequence of delta
functions
$$
\varkappa(t) := 
\sum_{j \in \mathbb{Z}} \sum\limits_{l=1}^n c_l \ \delta(t-(j+\alpha_l)),
$$
with fixed $0<\alpha_1 < \cdots < \alpha_n<1$ and real constants $c_j$ satisfying $c_1,  \ldots, c_n<1$ with $c_1 + \cdots +c_n =0$ (this ensures that the dynamical decoupling condition \fer{dyn.cond} is satisfied). Then we have  
\[
\int_0^t \varkappa(s) \, \mathrm{d}s  = \left\{
\begin{array}{ccl}
\sum_{j \in \mathbb{Z}} \sum\limits_{k=1}^n 
c_l \mathbf{1}[j+\alpha_l \in [0,t)]
&,& t> 0, \\
0 &,& t=0,\\
-\sum_{j \in \mathbb{Z}} \sum\limits_{l=1}^n c_l 
\mathbf{1}[j+\alpha_l \in [t,0)]
&,& t < 0,
\end{array} \right.
\]
and so $Q_a(t)
=\mathrm{e}^{-iH_{\mathrm{s}}\int_{0}^{t}\varkappa (s)\mathrm{d}s} q_{a
} \mathrm{e}^{iH_{\mathrm{s}}\int_{0}^{t}\varkappa (s)\mathrm{d}s}$ 
 is also a  $1$-periodic, piece-wise constant functions of the variable 
$t \in \mathbb{R}$. The Fourier coefficients become (see \fer{mmm1})
$$
Q_{k,a}:= \int_0^1 Q_a(t)\ \mathrm{e}^{-2\pi ikt} \ \mathrm{d}t
= -\frac{i}{2\pi  k}\left(\sum\limits_{l=1}^n
\mathrm{e}^{-2\pi i\alpha_l k} \ \delta Q_l
\right)
$$ 
if $k \in \mathbb{Z}\backslash \{0\}$, and $Q_{0,a}=0$ (because of the dynamical decoupling condition). Here,
$
\delta Q_l := \lim\limits_{\varepsilon \searrow 0}
(Q_a(t+\varepsilon)-Q_a(t-\varepsilon)).
$
Thus $\|Q_{k,a}\|$ is proportional to $1/|k|$ and  the decoherence rates to lowest order in $\lambda$ is
\begin{equation*}
\lambda^2\xi(T) \sim \lambda ^{2}\sum\limits_{k \in \mathbb{Z} \backslash \{0\}}\frac{1}{k^{2}}\, \left\{ | G_f(k/T+2)|^{2}+G_f(k/T-2)|^{2}\right\}.  \label{sim Facchi}
\end{equation*}
This behaviour of the decoherence rate was obtained in \cite{Facchi}, for a model describing one spin
interacting with a generic reservoir at inverse temperature $\beta$ (see equation
(110) in that reference). The analysis of \cite{Facchi} is formal, however, and based on functional integrals and Markov approximations.

\section{Proof of Theorem \protect\ref{mainthm}\label{section proof}}
\label{sectproof}

\subsection{The GNS\ representation}

Let $\omega _{\mathcal{S}}$ be a faithful state on $\mathcal{B}(\mathbb{C}%
^{d})$. Its GNS representation $(\mathfrak{H}_{\mathrm{s}},\pi _{\mathrm{s}%
},\Omega _{\mathrm{s}})$ is given explicitly as follows. The representation
Hilbert space is the linear space $\mathfrak{H}_{\mathrm{s}}=\mathcal{B}({%
\mathbb{C}}^{d})$ with inner product $\langle A,B\rangle _{\mathrm{s}}=%
\mathrm{Tr}(A^{\ast }B)$. The representation map $\pi _{\mathrm{s}}:%
\mathcal{B}(\mathbb{C}^{d})\rightarrow \mathcal{B}(\mathfrak{H}_{\mathrm{s}%
})$ is the left multiplication, $\pi _{\mathrm{s}}\left( A\right) =%
\underrightarrow{A}$, where for $A\in \mathcal{B}(\mathbb{C}^{d})$, we
define the left and right multiplication operators $\underrightarrow{A}$ and 
$\underleftarrow{A}$, acting on $\mathfrak{H}_{\mathrm{s}}$, as in \fer{left multi}. The system von Neumann algebra of observables (acting on the GNS Hilbert space) is 
\begin{equation}
{\mathfrak M}_\S= \underrightarrow{{\mathcal B}({\mathbb C}^d)} = \{\underrightarrow{A}\ :\ A\in\mathcal{B}(\mathbb{C}^{d})\}.
\label{underMS}
\end{equation}
The cyclic vector of the GNS representation of $\omega _{\mathcal{S}}$ is
given by the vector $\Omega _{\mathrm{s}}=\rho _{\mathrm{s}}^{1/2}\in 
\mathfrak{H}_{\mathrm{s}}$, where $\rho _{\mathrm{s}}\in \mathcal{B}(%
\mathbb{C}^{d})$ is the density matrix associated to $\omega _{\mathcal{S}}$%
. We have $\omega _{\mathcal{S}}\left( A\right) =\langle \Omega _{\mathrm{s}%
},\underrightarrow{A}\Omega _{\mathrm{s}}\rangle _{\mathrm{s}}$ for $A\in 
\mathcal{B}(\mathbb{C}^{d})$. The dynamics $\tau _{t}^{\mathrm{s}}$, generated by the derivation $\delta_{\rm s}=i[H_\S,\, \cdot\,]$ (see also (\ref{m101})), is implemented by the self-adjoint Liouville operator 
\begin{equation}
L_{\mathrm{s}}:=\left( \underrightarrow{H_{\mathrm{s}}}-\,\underleftarrow{H_{%
\mathrm{s}}}\right) =[H_{\mathrm{s}},\cdot ],  \label{liouvillean atom}
\end{equation}%
we have $\pi (\tau _{t}^{\mathrm{s}}(A))=\mathrm{e}^{itL_{\mathrm{s}}}\pi (A)%
\mathrm{e}^{-itL_{\mathrm{s}}}$. Note that $L_{\mathrm{s}}\Omega _{\mathrm{s}%
}=0$.

We now present the GNS representation of the infinitely extended free Fermi
reservoir at inverse temperature $\beta $. It is given by the Araki--Wyss
representation of the CAR algebra ${\mathcal{V}}_{\mathcal{R}}$ \cite%
{JaksicPillet, ArakiWyss}. The representation Hilbert space is 
\[
\mathfrak{H}_{\mathcal{R}}:={\mathcal{F}}_{-}(L^{2}(\mathbb{R}\times S^{2})),
\]%
the anti-symmetric Fock space over the one-particle space $L^{2}(\mathbb{R}%
\times S^{2})$ (equipped with the product of Lebesgue measure on $\mathbb{R}$
times the uniform measure on the two-sphere $S^{2}$). This representation is customarily referred to as \textit{Jaksic-Pillet glueing} and appeared for the first time in \cite{JP1}. The cyclic vector $\Omega _{\mathcal{R}}$ is the vacuum of ${\mathcal{F}}_{-}(L^2({\mathbb R}\times S^2))$. The representation map $\pi _{\mathcal{R}}$ maps an annihilation operator $%
a(f)\in {\mathcal{V}}_{\mathcal{R}}$, $f\in L^{2}({\mathbb{R}}^{3})$, to an
annihilation operator acting on $\mathfrak{H}_{\mathcal{R}}$, $\pi _{%
\mathcal{R}}(a(f))=a(g_{f})$, where $g_{f}\in L^{2}({\mathbb{R}}\times
S^{2}) $ is given by \fer{fg1}. We have $\omega _{\mathcal{R}}\left( A\right) =\left\langle \Omega _{%
\mathcal{R}},\pi _{\mathcal{R}}\left( A\right) \Omega _{\mathcal{R}%
}\right\rangle $ for all $A\in {\mathcal{V}}_{\mathcal{R}}$.

The Bogoliubov automorphism (\ref{definition automorphism reservoir}),
defined on ${\mathcal{V}}_{{\mathcal{R}}}$, extends to a $\ast $%
--auto\-mor\-phism group on the von Neumann algebra 
\begin{equation}
\mathfrak{M}_{\mathcal{R}}={\mathcal{V}}_{{\mathcal{R}}}^{\prime \prime }
\label{fieldvna}
\end{equation}
(weak closure of ${\mathcal{V}}_{{\mathcal{R}}}$). We denote the extension again by $\tau _{t}^{{\mathcal{R}}}$. The thermal state $\omega _{\mathcal{R}}$ is a $(\beta ,\tau_t^{{\mathcal{R}}%
})$--KMS state on $\mathfrak{M}_{{\mathcal{R}}}$. The dynamics is
implemented by the self-adjoint Liouville operator 
\begin{equation}
L_{\mathcal{R}}=\mathrm{d}\Gamma (p),
\end{equation}
the second quantization of the multiplication operator by the radial
variable $p\in\mathbb{R}$ of functions in $L^2({\mathbb{R}}\times S^2)$. We
have $\pi_{\mathcal{R}}(\tau_t^{\mathcal{R}}(A)) = \mathrm{e}^{i tL_{%
\mathcal{R}}}\pi_{\mathcal{R}}(A) \mathrm{e}^{-i tL_{\mathcal{R}}}$ for all $%
A\in{\mathcal{V}}_{{\mathcal{R}}}$, and $L_{\mathcal{R}}\Omega_{\mathcal{R}%
}=0$.

The GNS representation $(\mathfrak{H},\pi ,\Omega)$ of the initial state (%
\ref{initialstate}) is given by 
\begin{equation}
\mathfrak{H}:=\mathfrak{H}_{\mathrm{s}}\otimes \mathfrak{H}_{\mathcal{R}}\
,\quad \pi :=\pi _{\mathrm{s}}\otimes \pi _{\mathcal{R}}\ ,\quad \Omega
_{0}:=\Omega _{\mathrm{s}}\otimes \Omega _{\mathcal{R}}.
\label{omeganot}
\end{equation}
We denote by 
$$
\mathfrak{M}=\mathfrak{M}_\S \otimes \mathfrak{M}_{{\mathcal{R}}}
$$
the von Neumann algebra of observables of the joint system, see also \fer{underMS} and \fer{fieldvna}. The self-adjoint Liouville operator 
\begin{equation}
L=L_{\mathrm{s}}\otimes \mathbf{1}_{\mathfrak{H}_{\mathcal{R}}}+\mathbf{1}_{%
\mathfrak{H}_{\mathrm{s}}}\otimes L_{\mathcal{R}}=(\underrightarrow{H_{%
\mathrm{s}}}-\,\underleftarrow{H_{\mathrm{s}}})\otimes \mathbf{1}_{\mathfrak{%
H}_{\mathcal{R}}}+\mathbf{1}_{\mathfrak{H}_{\mathrm{s}}}\otimes \mathrm{d}
\Gamma (p),  \label{liouvillean free}
\end{equation}
defines the uncoupled Heisenberg dynamics $\mathrm{e}^{itL} A\mathrm{e}^{-itL}$ on $\mathfrak{M}$.

\subsection{Time-dependent Liouville operator}

As $\Omega _{\mathcal{R}}$ defines a KMS state on $\mathfrak{M}_{\mathcal{R}%
} $, the vector $\Omega _{\mathcal{R}}$ is cyclic and separating for $%
\mathfrak{M}_{\mathcal{R}}$. Let $\Delta _{\mathcal{R}}$ and $J_{\mathcal{R}%
} $ be the modular operator and conjugation of $\left( \mathfrak{M}_{%
\mathcal{R}},\Omega _{\mathcal{R}}\right) $ \cite{BR1}. Similarly, since $%
\omega _{\mathcal{S}}$ is faithful, $\Omega _{\mathrm{s}}=\rho _{\mathrm{s}%
}^{1/2}$ is cyclic and separating for the von Neumann algebra $\mathfrak{M}_{%
\mathrm{s}}$. Let $\Delta _{\mathrm{s}}$ and $J_{\mathrm{s}}$ be the associated modular
operator and conjugation. We set $J=J_{\mathrm{s}}\otimes J_{\mathcal{R}}$
and $\Delta =\Delta _{\mathrm{s}}\otimes \Delta _{\mathcal{R}}$ and define
the time-dependent Liouville operator, for $t\geq 0$, by 
\begin{eqnarray}
\mathcal{L}_{t} &=&L+V_{t},  \label{Lt} \\
V_{t} &=&W_{t}-J\Delta ^{1/2}W_{t}\Delta ^{-1/2}J,
\end{eqnarray}%
where 
\begin{equation}
W_{t}=\lambda \ \underrightarrow{Q(t)}\otimes \frac{1}{\sqrt{2}}\left(
a(g_{f})+a^{\ast }(g_{f})\right)\equiv \lambda Q(t)\otimes\Phi(g_f),  
\label{Wt}
\end{equation}
and 
\begin{eqnarray}
J\Delta ^{1/2}W_{t}\Delta ^{-1/2}J&=&\lambda \underleftarrow{\rho _{\mathrm{s}%
}^{-1/2}Q(t)\rho _{\mathrm{s}}^{1/2}}\otimes \frac{(-1)^{\mathrm{d}\Gamma (%
\mathbf{1})}}{\sqrt{2}}\left( a(i\mathrm{e}^{-\beta p}g_{f})+a^{\ast
}(ig_{f})\right)\nonumber\\
 &\equiv& \lambda \underleftarrow{\rho _{\mathrm{s}
}^{-1/2}Q(t)\rho _{\mathrm{s}}^{1/2}}\otimes\Phi'(g_f).  \label{Wt'}
\end{eqnarray}
Here, $\mathrm{d}\Gamma (\mathbf{1})$ is the second quantization of $\mathbf{%
1}$, i.e., the particle number operator acting on $\mathfrak{H}_{\mathcal{R}%
} $. 

\subsection{Implementation of the dynamics}

A family $\{U_{t,s}\}_{t\geq s}$ of bounded operators on $\mathfrak{H}$ is
called an evolution family if it satisfies (i) $U_{t,s}=U_{t,r}U_{r,s}$ for
all $t\geq r\geq s$ (cocycle, or Chapman--Kolmogorov property) and (ii) $%
U_{t,s}$ is a strongly continuous two--parameter family.

\begin{lemma}
\label{section extension} There is an evolution family $\{U_{t,s}\}_{t\geq
s}\subset \mathcal{B}\left( \mathfrak{H}\right) $ solving the following
non--autonomous evolution equations on $\mathrm{Dom}\left( L\right) $, 
\[
\forall t>s:\quad \partial _{t}U_{t,s}=i\mathcal{L}_{t}U_{t,s}\ ,\quad
\partial _{s}U_{t,s}=-iU_{t,s}\mathcal{L}_{s}\ ,\quad U_{s,s}:=\mathbf{1}.
\]%
For any $t\geq s$, $U_{t,s}$ possesses a bounded inverse $U_{t,s}^{-1}$.
Moreover, we have $U_{t,s}=U_{t+T,s+T}$ for all $t\geq s$.
\end{lemma}

A proof of this result is not difficult. For instance, one can follow the
ideas of \cite[Theorem X.69 and comments thereafter]{RS}.

\begin{satz}[Implementing the dynamics]
\label{corolario legal copy(1)} We have for all $t\geq s$, $A\in \mathfrak{M}$, 
\begin{equation}
U_{t,s}\Omega _{0}=\Omega _{0} \qquad \mbox{and} \qquad \pi\left(\tau
_{t,s}^{I}\left( A\right)\right) =U_{t,s} \pi(A) U_{t,s}^{-1},
\label{45.1}
\end{equation}
where $\tau_{t,s}^I$ is defined in \fer{m-1}. 
\end{satz}

\noindent \textit{Proof of Theorem \ref{corolario legal copy(1)}.\ } We show
that for $t>s$ 
\begin{equation}
\frac{\d}{\d t} \scalprod{\phi}{U_{t,s}^{-1}\pi(\tau_{t,s}^I(A)) U_{t,s}\psi} =0,
\label{m1}
\end{equation}
for all $\phi ,\psi \in \mathrm{Dom}(L_{\mathcal{R}})$ and all $A\in \mathrm{%
Dom}(\delta _{\mathrm{r}})$. It then follows ($t\downarrow s$) that $%
U_{t,s}^{-1}\pi (\tau _{t,s}^{I}(A))U_{t,s}=\pi (A)$ for all $A\in \mathrm{%
Dom}(\delta _{\mathrm{r}})$ and hence for all $A\in \mathfrak{M}$, which is
the statement to be proven. Note that $U_{t,s}\Omega _{0}=\Omega _{0}$
follows from ${\mathcal{L}_{t}}\Omega _{0}=0$. By using $\partial
_{t}U_{t,s}=i\mathcal{L}_{t}U_{t,s}$ and the ensuing equation $\partial
_{t}(U_{t,s}^{-1})^{\ast }=i\mathcal{L}_{t}^{\ast }(U_{t,s}^{-1})^{\ast }$, we obtain
\begin{eqnarray}
\frac{\d}{\d t} \scalprod{\phi}{U_{t,s}^{-1}\pi(\tau_{t,s}^I(A)) U_{t,s}\psi}&=&\left\langle {i\mathcal{L}_{t}^{\ast }(U_{t,s}^{-1})^{\ast }\phi },{\pi
(\tau _{t,s}^{I}(A))U_{t,s}\psi }\right\rangle\nonumber\\
 &&+\left\langle {(U_{t,s}^{-1})^{\ast }\phi },{\pi (\tau _{t,s}^{I}(A))i\mathcal{L}%
_{t}U_{t,s}\psi }\right\rangle  \notag \\
&&+\left\langle {(U_{t,s}^{-1})^{\ast }\phi },{\pi \big(%
\delta _{t}^{I}(\tau _{t,s}^{I}(A))\big)U_{t,s}\psi }\right\rangle .
\label{m2}
\end{eqnarray}%
The term $J\Delta ^{1/2}W_{t}\Delta ^{-1/2}J$ in $\mathcal{L}_{t}$ and the
corresponding part in the adjoint $\mathcal{L}_{t}^{\ast }$ cancel out in
the first two terms in (\ref{m2}), as they commute with $\pi (\tau
_{t,s}^{I}(A))$. Thus we can replace both $\mathcal{L}_{t}$ and $\mathcal{L}%
_{t}^{\ast }$ by $L_{\mathrm{s}}+L_{\mathcal{R}}+W_{t}$ in (\ref{m2}). The
contributions coming from $L_{\mathrm{s}}+W_{t}$ in the first two terms
cancel the commutator term in $\delta _{t}^{I}(\tau _{t,s}^{I}(A))=\delta _{%
\mathrm{r}}(\tau _{t,s}^{I}(A))+i[H_{\mathrm{s}}+\lambda Q(t)\otimes \Phi (%
\mathrm{f}),\tau _{t,s}^{I}(A)]$. It follows that 
\begin{eqnarray}
\frac{\mathrm{d}}{\mathrm{d} t}
\left\langle
\phi, U_{t,s}^{-1}\pi(\tau^I_{t,s} (A))
U_{t,s} \psi
\right\rangle &=&
  \left\langle
iL_{\mathcal{R}}( U_{t,s}^{-1})^*\phi,\pi(\tau^I_{t,s} (A))
U_{t,s} \psi
\right\rangle \label{m3} \\
&&+\left\langle {(U_{t,s}^{-1})^{\ast }\phi },{\pi (\tau _{t,s}^{I}(A))iL_{%
\mathcal{R}}U_{t,s}\psi }\right\rangle  \notag \\
&&+\left\langle {(U_{t,s}^{-1})^{\ast }\phi },{\pi \big(\delta _{\mathrm{r}%
}(\tau _{t,s}^{I}(A))\big)U_{t,s}\psi }\right\rangle . \notag  
\end{eqnarray}%
The last term in (\ref{m3}) equals 
\begin{eqnarray}
\lefteqn{\frac{\mathrm{d}}{\mathrm{d}\alpha }\Big|_{\alpha =0}\scalprod{%
(U_{t,s}^{-1})^*\phi}{\pi\big(\tau^\R_\alpha(\tau_{t,s}^I(A))\big)
U_{t,s}\psi}}  \notag \\
&=&
\frac{\mathrm{d}}{\mathrm{d}\alpha }\Big|_{\alpha =0}
\left\langle {(U_{t,s}^{-1})^{\ast }\phi 
},{\mathrm{e}^{i\alpha L_{\mathcal{R}}}\pi \big(\tau _{t,s}^{I}(A)\big)%
\mathrm{e}^{-i\alpha L_{\mathcal{R}}}U_{t,s}\psi }\right\rangle ,  \notag
\end{eqnarray}%
which is exactly the negative of the sum of the first two terms on the right
side in that equation. \hfill $\blacksquare $

\subsection{Howland Generator for $U_{t,s}\label{Howland}$}

The generator $\mathcal{G}$, acting on ${\mathfrak H}_{\rm per}=L^2({\mathbb T}_T,{\mathfrak H})$ and defined by (\ref{howgen}), is given explicitly by 
\begin{equation}
\mathcal{G}=  i\frac{\mathrm{d}}{\mathrm{d}t}+L+V_{\mathrm{per}}
  \label{xx0}
\end{equation}
There is a core of $\mathcal{G}$ whose elements are differentiable functions $t\mapsto f(t)$ with $f(t)\in \mathrm{Dom}(L)$. The operators on the right side are
understood as follows: 
\begin{eqnarray}
\left( \textstyle\frac{\mathrm{d}}{\mathrm{d}t}f\right) \left(
t\right) &=&\textstyle\frac{\mathrm{d}}{\mathrm{d}t}f\left( t\right)\nonumber\\
\left(Lf\right) \left( t\right) &=&L\left( f\left( t\right) \right)\nonumber\\
\left( V_{\mathrm{per}}f\right) \left( t\right) &=& [W_{t}-J\Delta ^{1/2}W_{t}\Delta
^{-1/2}J]\left( f\left( t\right) \right).
\label{XX2}
\end{eqnarray}
Recall the definition of the Fourier transform \fer{FT} and the notation introduced after it. The operator $\widehat{{\mathcal{G}}}=\mathfrak{F}{\mathcal{G\mathfrak{F}}}^{\ast }$ on $\ell ^{2}(\mathbb{Z},\mathfrak{H})$ is
\begin{equation}
\widehat{{\mathcal{G}}} =- \frac{2\pi }{T}k +L +\widehat{V}_{\mathrm{per}}.
\label{holand fourier}
\end{equation}
Its domain consists of $\hat f$ satisfying $\sum_{k\in%
\mathbb{Z}}k^2\|\hat{f}(k)\|^2<\infty$ and $f(k)\in\mathrm{Dom}(L)$. Here,
for $k\in \mathbb{Z}$, 
$$
( k\hat{f}) ( k) :=k\hat{f}( k) \mbox{\quad and\quad} ( L\hat{f}) ( k)
:=L( \hat{f}( k)).
$$
The operator $\widehat{V}_{\mathrm{per}}:=\mathfrak{F}V_{\mathrm{per}}%
\mathfrak{F}^{\ast }$ acts as a convolution,
\begin{equation}
[\widehat{V}_{\mathrm{per}}\hat{f}
](k)=\sum_{n\in\mathbb{Z}} \widehat{v}_{\mathrm{per}}(n-k)\hat{f}(n),
\label{XX1}
\end{equation}
with convolution kernel
\begin{equation}
\widehat{v}_{\mathrm{per}}(\ell)=\frac 1T\int_0^T\mathrm{e}^{-2\pi i\ell t/T}
\{W_t-J\Delta^{1/2} W_t\Delta^{-1/2}J\}\d t.
\label{XX3}
\end{equation}
The operators $W_t$ and $J\Delta^{1/2} W_t\Delta^{-1/2}J$ are given in \fer{Wt} and \fer{Wt'}, respectively.

\begin{lemma}
\label{lemmalongtime}
For any initial state $\omega _{\mathcal{S}}$ and all $A\in \mathcal{B}(%
\mathbb{C}^{d})$, 
\begin{eqnarray}
\lefteqn{
\left\vert \left\langle \Omega _{0},U_{\alpha ,0}\pi (A\otimes \mathbf{1}_{%
\mathfrak{H}_{\mathcal{R}}})\Omega _{0}\right\rangle_{{\mathfrak H}}
-\left\langle \widehat Z\Omega
_{0},\mathrm{e}^{i\alpha \widehat{{\mathcal{G}}}} \widehat Z(\left( \pi (A\otimes \mathbf{1}_{\mathfrak{H}_{\mathcal{R}}})\Omega _{0})\right) \right\rangle _{\widehat{\mathrm{per}}}\right\vert}\qquad\qquad\qquad\qquad\qquad\nonumber\\
&&\qquad\qquad\leq C(1+|\lambda |)\left\Vert A\right\Vert T,
\label{longtimebehavior}
\end{eqnarray}
where $C$ is some finite constant not depending on $\omega _{\mathcal{S}}$, $%
A$, $\lambda $, $T$, and $\alpha $.
\end{lemma}

\noindent \textit{Proof of Lemma \ref{lemmalongtime}.\ } By unitarity of the
Fourier transform, we can replace $\mathrm{e}^{i\alpha \widehat{\mathcal{G}}}$
by $\mathrm{e}^{i\alpha \mathcal{G}}$. We then have 
\begin{eqnarray}
\lefteqn{\left\vert \left\langle \Omega _{0},U_{\alpha ,0}\pi (A\otimes 
\mathbf{1}_{\mathfrak{H}_{\mathcal{R}}})\Omega _{0}\right\rangle_{{\mathfrak H}}
-\left\langle  Z\Omega _{0},\mathrm{e}^{i\alpha {\mathcal{G}}} Z(\left(
\pi (A\otimes \mathbf{1}_{\mathfrak{H}_{\mathcal{R}}})\Omega _{0})\right)
\right\rangle_{{\mathrm{per}}}\right\vert }  \notag \\
&=& \left\vert \left\langle Z\Omega _{0},U_{\alpha ,0}^{\rm per}Z(\pi(A\otimes 
\mathbf{1}_{\mathfrak{H}_{\mathcal{R}}})\Omega _{0})\right\rangle_{{\mathrm{per}}}
-\left\langle Z\Omega _{0},\mathrm{e}^{i\alpha {\mathcal{G}}} Z(\left(
\pi (A\otimes \mathbf{1}_{\mathfrak{H}_{\mathcal{R}}})\Omega _{0})\right)
\right\rangle_{{\mathrm{per}}}\right\vert \nonumber\\
&\leq &\frac{1}{T}\int_{0}^{T}\left\vert \left\langle \Omega _{0},(U_{\alpha
,0}-U_{t,t-\alpha })\pi (A\otimes \mathbf{1}_{\mathfrak{H}_{\mathcal{R}%
}})\Omega _{0}\right\rangle_{{\mathfrak H}} \right\vert \d t.  \label{m4}
\end{eqnarray}%
We show that (\ref{m4}) is small. For this, we want to shift the
time-indices of the $U_{t,t-\alpha }$ close to those of $U_{\alpha ,0}$ by
adding multiples of $T$ and using periodicity. The approximation is better
the smaller $T$ is. Given any $t\in \lbrack 0,T)$ and $\alpha \geq 0$, there
is an integer $n=n(t,\alpha )$ and an $\epsilon =\epsilon (t,\alpha )$ s.t. $%
n\geq 0$, $0\leq \epsilon <T$ and $t+nT=\alpha +\epsilon $. Since $%
U_{t,t-\alpha }=U_{t+nT,t+nT-\alpha }=U_{\alpha +\epsilon ,\epsilon }$, it
suffices to show that 
\begin{equation}
\left\vert \left\langle \Omega _{0},(U_{\alpha ,0}-U_{\alpha +\epsilon
,\epsilon })\pi (A\otimes \mathbf{1}_{\mathfrak{H}_{\mathcal{R}}})\Omega
_{0}\right\rangle_{{\mathfrak H}} \right\vert \leq C(1+|\lambda |)\Vert A\Vert \epsilon ,
\label{m5}
\end{equation}%
uniformly in $\alpha \geq 0$. Uniformity in $\alpha $ holds true because $U$
implements a norm-preserving map (dynamics) on the algebra $\mathfrak{M}$. We
have 
\begin{equation}
\left\langle \Omega _{0},(U_{\alpha ,0}-U_{\alpha +\epsilon ,\epsilon })\pi
(A\otimes \mathbf{1}_{\mathfrak{H}_{\mathcal{R}}})\Omega _{0}\right\rangle_{{\mathfrak H}}
=-\int_{0}^{\epsilon }\left\langle \Omega _{0},\partial _{s}U_{\alpha
+s,s}\pi (A\otimes \mathbf{1}_{\mathfrak{H}_{\mathcal{R}}})\Omega
_{0}\right\rangle_{{\mathfrak H}} \d s.  \label{m6}
\end{equation}%
Now $\partial _{s}U_{\alpha +s,s}=i\mathcal{L}_{\alpha +s}U_{\alpha
+s,s}-U_{\alpha +s,s}(i\mathcal{L}_{s})$, with $\mathcal{L}_{t}$ given in (%
\ref{Lt}). We analyze the term $-U_{\alpha +s,s}(i\mathcal{L}_{s})$, the
other one is dealt with similarly. We have 
\begin{eqnarray*}
\lefteqn{\left\langle \Omega _{0},U_{\alpha +s,s}\mathcal{L}_{s}\pi(A\otimes \mathbf{1}%
_{\mathfrak{H}_{\mathcal{R}}})\Omega _{0}\right\rangle_{{\mathfrak H}}} \\
&=&\left\langle \Omega _{0},U_{\alpha +s,s}\pi ([H_{\mathrm{s}}\otimes 
\mathbf{1}_{\mathfrak{H}_{\mathcal{R}}}+\lambda Q(s)\otimes \Phi (\mathrm{f}%
),A\otimes \mathbf{1}_{\mathfrak{H}_{\mathcal{R}}}])\Omega _{0}\right\rangle_{{\mathfrak H}}
\\
&=&\left\langle \Omega _{0},\pi \big(\tau _{\alpha +s,s}^{I}([H_{\mathrm{s}%
}\otimes \mathbf{1}_{\mathfrak{H}_{\mathcal{R}}}+\lambda Q(s)\otimes \Phi (%
\mathrm{f}),A\otimes \mathbf{1}_{\mathfrak{H}_{\mathcal{R}}}])\big)\Omega
_{0}\right\rangle_{{\mathfrak H}},
\end{eqnarray*}
which has modulus bounded above by 
$$
\big(2\|H_{\mathrm{s}}\| +2|\lambda|\ \|Q\|\ \|f\|\big) \|A\|\leq C(1+|\lambda |)\Vert A\Vert,
$$
uniformly in $\alpha ,s$. Using this bound in (\ref{m6}) we obtain (\ref{m5}).\hfill $%
\blacksquare $

\medskip The next result implies our main result, Theorem \ref{mainthm} (see
details given after Theorem \ref{section extension copy(4)}). We give the
proof of Theorem \ref{section extension copy(4)} in the next section.

Denote by $\widehat{{\mathcal{G}}}_{0}$ the Howland generator for $\lambda =0$
(uncoupled system).

\begin{satz}
\label{section extension copy(4)} Suppose the $T$--periodic control term $H_{%
\mathrm{c}}$ satisfies the dynamical decoupling condition (\ref{dyn.cond}).
Then 
\begin{eqnarray*}
\lefteqn{
\left\vert \left\langle\widehat Z \Omega _{0},(\mathrm{e}^{i\alpha \widehat{{\mathcal{G}}}_{0}}-\mathrm{e}^{i\alpha \widehat{{\mathcal{G}}}})\widehat Z(\left( \pi (A\otimes 
\mathbf{1}_{\mathfrak{H}_{\mathcal{R}}})\Omega _{0})\right) \right\rangle_{\widehat{\mathrm{per}}}
\right\vert}\qquad\qquad\\
&& \leq 
C\left\Vert A\right\Vert \Big( |\lambda |+(D|\lambda|+1)T+\mathrm{e}^{c\alpha |\lambda |T}-1\Big)
\end{eqnarray*}
for all $\alpha \geq 0$ and all $A\in \mathcal{B}(\mathbb{C}^{d})$, and
where $0<c,C<\infty $ are constants not depending on $\alpha ,\lambda ,T,H_{%
\mathrm{c}}$, nor on $\omega_{\mathcal S}$ in the initial state $\omega _{0}=\omega _{\mathcal{S}%
}\otimes \omega _{{\mathcal{R}}}$. Here, $D$ is given in (\ref{mD}).
\end{satz}

\noindent \textbf{Proof of Theorem \ref{mainthm}. }\ Since the left side of
the inequality in Theorem \ref{section extension copy(4)} is bounded above
by $C\Vert A\Vert $ uniformly in $\alpha \geq 0$, we can replace $\mathrm{e}%
^{c\alpha |\lambda |T}-1$ by $\min \{1,\mathrm{e}^{c\alpha |\lambda |T}-1\}$%
. Moreover, since we have $\min \{1,\mathrm{e}^{a}-1\}\leq 2(1-\mathrm{e}%
^{-a})$ for all $a\geq 0$, the upper bound in Theorem \ref{section extension
copy(4)} can be replaced by 
$C\Vert A\Vert \{|\lambda |+(D|\lambda|+1)T+1-\mathrm{e}%
^{-c\alpha |\lambda |T}\}$. Next, note that 
\[
\mathrm{e}^{i\alpha \widehat{{\mathcal{G}}}_{0}}\widehat Z(\pi (A\otimes \mathbf{1}_{%
\mathfrak{H}_{\mathcal{R}}})\Omega _{0})=\widehat Z\big(\pi ( \mathrm{e}^{i\alpha H_{%
\mathrm{s}}}A\mathrm{e}^{-i\alpha H_{\mathrm{s}}}\otimes \mathbf{1}_{\mathfrak{H}_{\mathcal{R}}}) \Omega _{0}\big).
\]%
Combining Lemma \ref{lemmalongtime} and Theorem \ref{section extension
copy(4)} thus gives 
\begin{eqnarray}
\lefteqn{
\left\vert \omega _{0}(\tau _{\alpha ,0}^{I}(A\otimes \mathbf{1}_{\mathfrak{H%
}_{\mathcal{R}}}))-\omega _{\mathcal{S}}(\mathrm{e}^{i\alpha H_{\mathrm{s}}}A%
\mathrm{e}^{-i\alpha H_{\mathrm{s}}})\right\vert}\nonumber\\
&& \leq C\Vert A\Vert
\big(|\lambda |+(D|\lambda|+1)T+1-\mathrm{e}^{-c\alpha |\lambda |T}\big)  
\label{m-2}
\end{eqnarray}
Since $\tau _{\alpha ,0}^{I}(A\otimes \mathbf{1}_{\mathfrak{H}_{\mathcal{R}%
}})=V_{c}(\alpha )^{\ast }\tau _{\alpha ,0}(A\otimes \mathbf{1}_{\mathfrak{H}%
_{\mathcal{R}}})V_{c}(\alpha )$ (see (\ref{m-1})), and since the bound (\ref%
{m-2}) in uniform in the initial state $\omega _{\mathcal{S}}$, we obtain
the assertion (\ref{m-3}). \hfill $\blacksquare $

\subsection{Resonances of Howland Generator, Proof of Theorem \protect\ref{section extension copy(4)}}

\label{sectionspectral deformation}

Now we perform an analytic deformation of the Howland operator $\widehat{%
\mathcal{G}}=\mathfrak{F}{\mathcal{G\mathfrak{F}}}^{\ast }$ (\ref{holand
fourier}) acting on $\widehat{\mathfrak{H}}_{\mathrm{per}}=\ell ^{2}(\mathbb{%
Z},\mathfrak{H})$. For all $\theta \in \mathbb{C}$, define 
\begin{equation}
\widehat{\mathcal{G}}_{0}(\theta ):=- \frac{2\pi }{T}k+L+\theta \widehat{N%
},  
\label{definition G non perturbe}
\end{equation}%
where $\widehat{N}(\hat{f})(k):=(\mathbf{1}_{\mathfrak{H}_{\mathrm{s}}}\otimes 
\mathrm{d}\Gamma (\mathbf{1}))\hat{f}(k)$ for all integers $k$. For $\theta $
with $\mathrm{\func{Im}}\theta>0$, $\widehat{\mathcal{G}}_{0}(\theta )$ is a
normal operator with spectrum contained in the closed upper complex
half--plane  ${\mathbb C}^+$ and domain $\mathrm{Dom}(\widehat{\mathcal{G}}_{0}(\theta ))=%
\mathrm{Dom}(\widehat{\mathcal{G}}_{0})\cap \mathrm{Dom}(\widehat{N})$. In
particular, $\widehat{\mathcal{G}}_{0}(\theta )$ is the generator of a
strongly continuous contraction semigroup for such $\theta$.

Recall that $\widehat V_{\rm per}$ is defined as ${\mathfrak F}V_{\rm per}{\mathfrak F}^*$, see \fer{holand fourier}. This operator acts as a convolution, \fer{XX1}, having the convolution kernel $\widehat v_{\rm per}(\ell)$ given in \fer{XX3}. We define the spectrally deformed kernel by 
\begin{equation}
\widehat v_{\rm per}(\ell;\theta)=\frac 1T\int_0^T\mathrm{e}^{-2\pi i\ell t/T}
\{W_t(\theta)-(J\Delta^{1/2} W_t\Delta^{-1/2}J)(\theta)\}\d t,
\label{defokernel}
\end{equation}
where (see \fer{Wt}, \fer{Wt'})
\begin{eqnarray}
W_{t}(\theta)&=&\lambda \ \underrightarrow{Q(t)}\otimes \frac{1}{\sqrt{2}}\left(
a(g_{f,\bar\theta})+a^{\ast }(g_{f,\theta})\right)\nonumber\\
&\equiv& \lambda \ \underrightarrow{Q(t)}\otimes\Phi_\theta(g_f)  \label{Wtdefo}\\
(J\Delta ^{1/2}W_{t}\Delta ^{-1/2}J)(\theta) &=&\lambda \underleftarrow{\rho _{\mathrm{s}
}^{-1/2}Q(t)\rho _{\mathrm{s}}^{1/2}}\nonumber\\
&& \otimes \frac{(-1)^{\mathrm{d}\Gamma (\mathbf{1})}}{\sqrt{2}}\left( a(i\mathrm{e}^{-\beta(p+\bar\theta)}g_{f,\bar{\theta}})+a^{\ast
}(ig_{f,\theta})\right)\nonumber\\
&\equiv&   \lambda \underleftarrow{\rho _{\mathrm{s}
}^{-1/2}Q(t)\rho _{\mathrm{s}}^{1/2}} \otimes\Phi'_\theta(g_f).
\label{Wt'defo}
\end{eqnarray}
Here, 
\begin{equation}
g_{f,\theta}(p, \vth) := g_{f}(p+\theta, \vth).
\label{gftheta}
\end{equation}
We have put complex conjugates on $\theta$ in \fer{Wtdefo} and \fer{Wt'defo} in an appropriate way, so that $\theta\mapsto \widehat v_{\rm per}(\ell;\theta)$ is analytic in $\theta$, for $0<\mathrm{\func{Im}}\theta <r_{\max }$. Combining \fer{defokernel}, \fer{Wtdefo} and \fer{Wt'defo} we have the representation
\begin{equation}
\widehat v_{\rm per}(\ell;\theta) = \lambda  \underrightarrow{\widehat Q(\ell)}\otimes\Phi_\theta(g_f) +  \lambda \underleftarrow{\rho _{\mathrm{s}
}^{-1/2}\widehat Q(\ell)\rho _{\mathrm{s}}^{1/2}} \otimes\Phi'_\theta(g_f),
\label{vhatagain}
\end{equation}
where $\widehat Q(\ell)=\frac1T\int_0^t\e^{-2\pi i\ell t/T}Q(t)\d t$, see \fer{FT}. 

Let $\widehat V_{\rm per}(\theta)$ be the operator on $\widehat{\frak H}_{{\rm per}}$ given by the convolution with kernel ${\widehat v}_{\rm per}(\ell;\theta)$. Then $\theta \mapsto {\widehat V}_{\rm per}(\theta)$ is (bounded operator valued) analytic in $\theta$, for $0<\mathrm{\func{Im}}\theta <r_{\max }$, and we have the bound 
\begin{align} \label{eq-VB.5} 
\max_{\theta: 0\leq {\rm Im}\theta\leq r_{\rm max}} \big\| \hV_{\rm per} (\theta)\big\| \ \leq \ C \, |\lambda| , 
\end{align}
for some $C< \infty$ (see Assumption (A1) before (\ref{m34})). Since $\widehat{V}_{\mathrm{per}}\left( \theta \right)$ is bounded for any $\theta \in 
\mathbb{C}$ with $0<\mathrm{\func{Im}}\theta <r_{\max}$, the
deformed Howland generator 
\begin{equation}
\widehat{\mathcal{G}}(\theta ):=\widehat{\mathcal{G}}_{0}(\theta )+\widehat{V}_{%
\mathrm{per}}\left( \theta \right)   
\label{Howland deformed}
\end{equation}
is the generator of a strongly continuous semigroup, for each such $\theta $.
It is easy to see that $\Vert \mathrm{e}^{i\alpha \widehat{\mathcal{G}}%
_{0}(\theta )}\Vert =1$ and therefore \cite[Chapter III, 1.3]{EN}, 
\begin{equation}
\Vert \mathrm{e}^{i\alpha \widehat{\mathcal{G}}(\theta )}\Vert \leq \mathrm{e}%
^{\alpha \Vert \widehat{V}_{\mathrm{per}}(\theta )\Vert }.  
\label{m7}
\end{equation}
The subspace 
\[
\mathfrak{\widehat{D}}_{0}:=\left\{ \hat{f}:\;\hat{f}(\mathbb{Z})\subset \mathrm{%
Dom}(L)\;\text{and }\hat{f}(k)\neq 0,\text{ only for finitely many }k\in \mathbb{Z}%
\right\} 
\]%
is a core of $\widehat{\mathcal{G}}_{0}(\theta )$ for all $\theta \in 
\mathbb{C}$ with $0\leq \mathrm{\func{Im}}\theta $. $\mathfrak{\widehat{D}}_{0}$
is also a core of $\widehat{\mathcal{G}}(\theta )$ for such $\theta $, since 
$\widehat{V}_{\mathrm{per}}\left( \theta \right) $ is a bounded operator.

\begin{satz}[Deformation invariance of evolution]
\label{thminv} 
For all $\alpha \geq 0$, $\widehat{\psi}_{1},\widehat{\psi}_{2}\in \widehat Z (\mathfrak{H}_{\mathrm{s}}\otimes \Omega _{\mathcal{R}}) \subset 
\widehat{\mathfrak{H}}_{\mathrm{per}}$, and all $\theta \in \mathbb{C}$ with $%
\mathrm{\func{Im}}\theta \in \lbrack 0,r_{\max })$, we have 
\begin{equation}
\langle \widehat{\psi}_{1},\mathrm{e}^{i\alpha \widehat{\mathcal{G}}}\widehat{\psi}%
_{2}\rangle _{\widehat{\mathrm{per}}}=\langle \widehat{\psi}_{1},\mathrm{e}^{i\alpha 
\widehat{\mathcal{G}}(\theta )}\widehat{\psi}_{2}\rangle _{\widehat{\mathrm{per}}}.
\end{equation}
\end{satz}

\noindent \textit{Proof of Theorem \ref{thminv}.\ } We show that 
\begin{equation}
\langle \widehat{\psi}_{1},\mathrm{e}^{i\alpha\widehat{\mathcal{G}}(\theta ) }%
\widehat{\psi}_{2}\rangle _{\widehat{\mathrm{per}}}=\langle \widehat{\psi}_{1},\mathrm{e}%
^{i\alpha\widehat{\mathcal{G}}(\theta ^{\prime }) }\widehat{\psi}_{2}\rangle _{\widehat{\mathrm{per}}}  
\label{m8}
\end{equation}
for all $\theta ,\theta ^{\prime }\in \mathbb{C}$ with $\mathrm{\func{Im}}%
\theta ,\mathrm{\func{Im}}\theta ^{\prime }\in \lbrack 0,\mathrm{r}_{\max })$%
. The result follows since $\widehat{\mathcal{G}}=\widehat{\mathcal{G}}(0)$.

Note that $\widehat{\mathcal{G}}(\theta )\rightarrow \widehat{\mathcal{G}}$
strongly on $\mathfrak{\widehat{D}}_{0}$, as $\theta \rightarrow 0$ in the upper
complex half plane. It follows from the first Trotter-Kato approximation
theorem \cite[Chapter III, 4.8]{EN} that for all $\hat{f}\in \widehat{%
\mathfrak{H}}_{\mathrm{per}}$,

\begin{itemize}
\item[(a)] $\mathrm{e}^{i\alpha \widehat{\mathcal{G}}}\hat{f}=\lim_{\theta
\rightarrow 0} \mathrm{e}^{i\alpha \widehat{\mathcal{G}}(\theta )}\hat{f}$

\item[(b)] $(\zeta -\widehat{\mathcal{G}})^{-1}\hat{f}=\lim_{\theta
\rightarrow 0} (\zeta -\widehat{\mathcal{G}}(\theta ))^{-1}\hat{f}$, for $\zeta\in\mathbb{C}$ with $\mathrm{\func{Im}}\zeta <- \|\widehat{V}_{\mathrm{per}%
}\|$ (the contraction constant for $\theta=0$, see also (\ref{m7})).
\end{itemize}
The integral representation $(\zeta -\widehat{\mathcal{G}}(\theta ))^{-1}%
\hat{f}=-i\int_{0}^{\infty }\mathrm{e}^{i\alpha \widehat{\mathcal{G}}(\theta )}%
\mathrm{e}^{i\zeta \alpha }\hat{f}\ \mathrm{d}\alpha $ is valid for all $%
\zeta \in \mathbb{C}$ with $\mathrm{\func{Im}}\zeta <-\Vert \widehat{V}_{\mathrm{%
per}}(\theta )\Vert $ \cite[Chapter II, 1.10]{EN}. This representation,
together with the injectivity of the Laplace transform implies that in order
to prove (\ref{m8}), we only need to show 
\begin{equation}
\langle \widehat{\psi}_{1},(\zeta -\widehat{\mathcal{G}}(\theta ))^{-1}\widehat{\psi}_{2}\rangle_{\widehat{\mathrm{per}}}=\langle \widehat{\psi}_{1},(\zeta -\widehat{\mathcal{%
G}}(\theta ^{\prime }))^{-1}\widehat{\psi}_{2}\rangle _{\widehat{\mathrm{per}}}
\label{assertion}
\end{equation}
for any $\widehat{\psi}_{1},\widehat{\psi}_{2}\in \mathfrak{F}\left( \mathfrak{H}_{%
\mathrm{s}}\otimes \Omega _{\mathcal{R}}\right) \subset \widehat{\mathfrak{H}%
}_{\mathrm{per}}$, all $\theta ,\theta ^{\prime }\in \mathbb{C}$ with $%
\mathrm{\func{Im}}\theta ,\mathrm{\func{Im}}\theta ^{\prime }\in \lbrack 0,%
\mathrm{r}_{\max })$ and every $\zeta \in \mathbb{C}$ with imaginary part
sufficiently large and negative.

Let $\theta \in \mathbb{R}$. Define the translation operator $u(\theta )$ on 
$L^2({\mathbb{R}}^3)$ by $\left( u(\theta )f\right) \left( p,\vartheta
\right) =f\left( p+\theta ,\vartheta \right)$ and lift its action to $%
\widehat{\mathfrak{H}}_{\mathrm{per}}$ by setting 
\[
\forall \hat{f}\in \widehat{\mathfrak{H}}_{\mathrm{per}},\ k\in \mathbb{Z}%
:\quad (U(\theta )\hat{f})(k):=\left( \mathbf{1}_{\mathfrak{H}_{\mathrm{s}%
}}\otimes \Gamma (u(\theta ))\right) (\hat{f}(k)),
\]%
where $\Gamma (u(\theta ))$ is the second quantization of $u(\theta )$. We
have $U(\theta )=U(-\theta )^{\ast }$ for all $\theta \in \mathbb{R}$.
Observe that $(\zeta -\widehat{\mathcal{G}}(\theta ))^{-1}=U(\theta )(\zeta -%
\widehat{\mathcal{G}}(0))^{-1}U(\theta )^{\ast }$ and $U(\theta )\widehat{\psi}=%
\widehat{\psi}$ for $\theta \in \mathbb{R}$ and $\widehat{\psi}\in \widehat Z\left( \mathfrak{H}_{\mathrm{s}}\otimes \Omega_{\mathcal{R}}\right)$. It
follows that the function 
\[
g(\theta) :=\langle \widehat{\psi}_{1},(\zeta -\widehat{\mathcal{G}}%
(\theta ))^{-1}\widehat{\psi}_{2}\rangle_{\widehat{\mathrm{per}}}
\]%
is constant on $\mathbb{R}$, i.e., $g\left( \theta \right) =g\left( 0\right) 
$ for all $\theta \in \mathbb{R}$. The family $\{\widehat{\mathcal{G}}%
(\theta )\}_{\theta \in \mathbb{R}+i(0,r_{\max })}$ of closed
operators is of type A. Take $\zeta$ with $\mathrm{\func{Im}} \zeta
<-\sup_{\theta\in{\mathbb{R}}+i(0,r_{\mathrm{max}})}\|\widehat{V}_{\mathrm{per}%
}(\theta)\|$. Then $\theta\mapsto g(\theta)$ is analytic for all $\theta\in{%
\mathbb{R}}+i(0,r_{\mathrm{max}})$. By (b) above, the function is continuous
as $\theta$ approaches the real axis from above. Using the Schwarz reflection
principle, we deduce that $g$ must be constant on all of $\mathbb{R}+i[0,r_{\max })$. \hfill $\blacksquare$

\bigskip

Theorem \ref{thminv} shows that the evolution of the system is expressed by
the $C_{0}$-semigroup $\{\mathrm{e}^{i\alpha \widehat{\mathcal{G}}%
(ir)}\}_{\alpha \geq 0}$ at any fixed $r\in (8\Vert H_{\mathrm{s}}\Vert
,r_{\max })$. See again condition (A2). For $\lambda =0$, the spectrum of
the normal operator $\widehat{\mathcal{G}}_{0}(ir)$ is
\[
\sigma (\widehat{\mathcal{G}}_{0}(ir))=\sigma _{\mathrm{d}}(\widehat{
\mathcal{G}}_{0}(ir))\cup \big\{ {\mathbb R}+ir{\mathbb N}\big\},
\]
where
\[
\sigma _{\mathrm{d}}(\widehat{\mathcal{G}}_{0}(ir))=\frac{2\pi }{T}\mathbb{%
Z+}\sigma ([H_{\mathrm{s}},\cdot ])
\]
is the set of discrete eigenvalues of $\widehat{\mathcal{G}}_{0}(ir)$. The
spectrum of $[H_{\mathrm{s}},\cdot ]$ consists of all differences of eigenvalues of $H_{\mathrm{s}}$ (the so-called Bohr energies). The spectral
deformation separates the discrete and continuous spectrum by a distance $r$%
. This property allows us to apply standard analytic perturbation theory to
follow the eigenvalues under perturbation ($\lambda\neq 0$).

Let $\gamma _{k,\varepsilon }$, $k\in \mathbb{Z}$, be the positively
oriented circle with center $\frac{2\pi}{T}k$ and radius $\varepsilon$ satisfying $2\|H_\S\|<\varepsilon<\min\{\frac{\pi}{T},\frac{r}{2}\}$. We want the circles $\gamma _{k,\varepsilon }$ to contain exactly the eigenvalues of $\widehat{%
\mathcal{G}}(ir)$ bifurcating (for $\lambda \neq 0$) out of the eigenvalues $%
2\pi \ell /T+\sigma ([H_{\mathrm{s}},\cdot ])$ for $\ell =k$ (no
overlapping resonances). Therefore we impose the condition $\pi /T>2\Vert H_{%
\mathrm{s}}\Vert$, or, condition (A1). The projection valued maps 
\begin{equation}
\lambda \mapsto \mathcal{P}_{k}(\lambda ):=\frac{1}{2\pi i}%
\oint\limits_{\gamma _{k,\varepsilon }}(\zeta -\widehat{\mathcal{G}}%
(ir))^{-1}\mathrm{d}\zeta 
\label{61.5}
\end{equation}
are analytic in some ball of radius $c>0$ and center $0$ in the complex
plane, with $c$ independent of $k\in \mathbb{Z}$. We have
\begin{equation}
\mathcal{P}_{0}(0)=\mathbf{1}_{\mathfrak{H}_{\mathrm{s}}}\otimes |\widehat Z(\Omega _{\mathcal{R}})\rangle \langle \widehat Z(\Omega _{\mathcal{R}})|.
\label{m13'}
\end{equation}
To ease the readability of the equations to follow, we define
\begin{equation}
\widehat\Psi(A) = \widehat Z(\pi(A\otimes\mathbf{1}_{{\mathfrak H}_{\mathcal{R}}})\Omega_0),
\label{notation1}
\end{equation}
so that 
\begin{equation}
\left\langle\widehat Z \Omega _{0},\mathrm{e}^{i\alpha \widehat{\mathcal{G}}%
}\widehat Z(\left( \pi (A\otimes \mathbf{1}_{\mathfrak{H}_{\mathcal{R}}})\Omega
_{0}\right)) \right\rangle_{\widehat{\rm per}} =\scalprod{\widehat\Psi(\mathbf{1})}{\mathrm{e}^{i\alpha \widehat{\mathcal{G}}%
}\widehat\Psi(A)}_{{\widehat{\rm per}}}.
\label{notation2}
\end{equation}
By the Laplace inversion formula \cite[Chapter III, Corollary 5.15]{EN}, 
\begin{eqnarray*}
\lefteqn{\scalprod{\widehat\Psi(\mathbf{1})}{\mathrm{e}^{i\alpha \widehat{\mathcal{G}}}\widehat\Psi(A)}_{{\widehat{\rm per}}}} \\
&=&\lim_{\ell \rightarrow \infty }\frac{1}{2\pi i}\int_{\tilde{\gamma}_{\ell
}}\mathrm{e}^{i\alpha z}\scalprod{\widehat\Psi(\mathbf{1})}{(z-\widehat{\mathcal{G}}(ir))^{-1}\widehat\Psi(A)}_{{\widehat{\rm per}}} \mathrm{d}z,
\end{eqnarray*}%
where $\tilde{\gamma}_{\ell }$ is the straight contour from $-\ell -i$ to $\ell -i$. Deforming contours we obtain 
\begin{eqnarray}
\lefteqn{\scalprod{\widehat\Psi(\mathbf{1})}{\mathrm{e}^{i\alpha \widehat{\mathcal{G}}}\widehat\Psi(A)}_{{\widehat{\rm per}}}} \nonumber \\
&=&\lim_{\ell \rightarrow \infty }\Big[\frac{1}{2\pi i}\int_{\gamma _{\ell }}%
\mathrm{e}^{i\alpha z}\scalprod{\widehat\Psi(\mathbf{1})}{(z-\widehat{\mathcal{G}}(ir))^{-1}\widehat\Psi(A)}_{{\widehat{\rm per}}} \mathrm{d}z   \notag \\
&&+\sum\limits_{\substack{ k\in \mathbb{Z}, \\ |k\frac{2\pi }{T}|<\ell }}%
\scalprod{\widehat\Psi({\mathbf{1}})}{\mathrm{e}^{i\alpha \widehat{{\mathcal{G}}}(ir)}\mathcal{P}_{k}(\lambda )\widehat\Psi(A)}_{{\widehat{\rm per}}} \Big],  
\label{m10}
\end{eqnarray}
where $\gamma _{\ell }$ is the straight contour from $-\ell +ir/2$ to $\ell
+ir/2$. Indeed, observe that, on the domain of $\widehat{\mathcal{G}}(ir)$,  
\begin{equation}
(z-\widehat{\mathcal{G}}(ir))^{-1}=\frac{1}{z}\big[(z-\widehat{\mathcal{G}}%
(ir))^{-1}\widehat{\mathcal{G}}(ir)+\mathbf{1}\big]  \label{Identite super}
\end{equation}%
and thus $(z-\widehat{\mathcal{G}}(ir))^{-1}\widehat\Psi(A)=\mathcal{O}(\ell ^{-1})$ on the
straight vertical paths joining $z=\pm \ell -i$ and $z=\pm \ell +ir/2$.

\begin{lemma}[ Bounds for $\mathcal{P}_{k}(\protect\lambda )$]
\label{lemmabounds} For some $C<\infty $, all $A\in \mathcal{B}(\mathbb{C}%
^{d})$, and all $k\in \mathbb{Z}\backslash \{0\}$, we have 
\[
\left\Vert \mathcal{P}_{k}(\lambda )\widehat \Psi(A) \right\Vert_{{\widehat{\rm per}}} \leq C\left\Vert A\right\Vert
T^{2}k^{-2}\left( 1+|\lambda |\max_{0\leq t\leq T}\left\Vert H_{\mathrm{c}%
}(t)\right\Vert \right) .
\]
\end{lemma}

\noindent \textit{Proof of Lemma \ref{lemmabounds}.\ } Applying (\ref%
{Identite super}) twice and observing that $\frac{1}{\zeta }$, $\frac{1}{%
\zeta ^{2}}$ are analytic away from $\zeta =0$ we get 
\begin{eqnarray}
\mathcal{P}_{k}(\lambda )\widehat\Psi(A)&=&\frac{1}{2\pi i}\oint\limits_{\gamma
_{k,\varepsilon }}\frac{1}{\zeta }(\zeta -\widehat{\mathcal{G}}(ir))^{-1}%
\widehat{\mathcal{G}}(ir)\widehat\Psi(A) \mathrm{d}\zeta   \notag \\
&=&\frac{1}{2\pi i}\oint\limits_{\gamma _{k,\varepsilon }}%
\frac{1}{\zeta ^{2}}(\zeta -\widehat{\mathcal{G}}(ir))^{-1}\widehat{\mathcal{%
G}}(ir)^{2}\widehat\Psi(A) \mathrm{d}\zeta .  
\label{doublestarr}
\end{eqnarray}
We obtain now the bound $\widehat{\mathcal{G}}(ir)^{2}\widehat\Psi(A) =\mathcal{O}(T^{0}\lambda
^{0}+\lambda \max_{0\leq t\leq T}\left\Vert H_{\mathrm{c}}(t)\right\Vert )$,
which, together with $\zeta ^{-2}=\mathcal{O}(T^{2}/k^{2})$, implies the
bound of Lemma \ref{lemmabounds}. Observe that, since $L_\R\widehat\Psi(A)=0$, $\widehat N\widehat\Psi(A)=0$ and $k\widehat\Psi(A)=0$, we have 
\begin{eqnarray*}
\widehat{\mathcal{G}}(ir)^{2}\widehat \Psi(A) &=&\big(\widehat{\mathcal{G}}_{0}(ir)+\widehat{V}_{\mathrm{per}}\left( ir\right) \big)(L_{\mathrm{s}}+\widehat{V}_{\mathrm{per}}\left( ir\right) )\widehat\Psi(A) \\
&=&\big(L_{\mathrm{s}}^{2}+\widehat{V}_{\mathrm{per}}\left( ir\right) ^{2}+\widehat{V}_{%
\mathrm{per}}\left( ir\right) L_{\mathrm{s}}+\widehat{\mathcal{G}}_{0}(ir)\widehat{V}_{\mathrm{per}}\left( ir\right) \big)\widehat\Psi(A),
\end{eqnarray*}
from which it follows that 
$$
\|\widehat{\cal G}(ir)^2\widehat\Psi(A)\|_{\widehat{\rm per}} \leq C\|A\| +\|\widehat{\cal G}_0(ir)\widehat V_{\rm per}(ir)\widehat\Psi(A)\|_{\widehat{\rm per}}. 
$$
To prove the lemma it suffices thus to show that 
\begin{equation}
\|\widehat{\mathcal{G}}_{0}(ir)\widehat{V}_{\mathrm{per}}\left( ir\right) \widehat\Psi(A)\|_{\widehat{\rm per}}\leq C\|A\|\, |\lambda| \big( 1+|\lambda|\max_{0\leq t\leq T}\|H_{\rm c}(t)\|\big).  
\label{m9}
\end{equation}
We have 
\begin{equation}
\widehat{\mathcal{G}}_{0}(ir)\widehat{V}_{\mathrm{per}}\left( ir\right)
\widehat\Psi(A)=(L_{\rm s}+L_\R+ir-2\pi k/T)\widehat{V}_{\mathrm{per}}\left( ir\right) \widehat\Psi(A).
\label{onebound}
\end{equation}
We use here that the vector $\widehat V_{\rm per}(ir)\widehat\Psi(A)$ is in the sector $\widehat N=1$, since $\widehat V_{\rm per}(ir)$ is linear in creation and annihilation operators and $\widehat\Psi(A)$ is proportional to the vacuum, see \fer{notation1} and \fer{omeganot}. The summand with $L_{\mathrm{s}}+ir$ in \fer{onebound} is bounded above by $C|\lambda|\,\|A\|$ (uniformly in $T$). Next, $\widehat V_{\rm per}(ir)= \widehat V_{{\rm per},1}(ir)+\widehat V_{{\rm per},2}(ir)$ is the sum of two terms, each one given as the convolution operator with kernel given by one of the summands in \fer{vhatagain}. We treat 
$$
\widehat V_{{\rm per},1}(ir) = \lambda {\frak F}Q(t){\frak F}^*\otimes\Phi_{ir}(g_f),
$$
the other one is estimated in the same fashion. We have 
$$
\|L_\R\widehat V_{{\rm per},1}(ir)\widehat \Psi(A)\|\leq |\lambda|\, \|Q\|\, \|A\|\, \|L_\R\Phi_{ir}(g_f)\Omega_\R\|\leq C|\lambda|\,\|A\|,
$$
due to condition (A2) and since $\|{\frak F}Q(t){\frak F}^*\|=\|Q(t)\|=\|Q\|$. Consider now the term containing the factor $-2\pi k/T$ in \fer{onebound},
$$
\| 2\pi kT^{-1}\widehat{V}_{\mathrm{per}}(ir)\| \leq C|\lambda |\, \Vert 2\pi kT^{-1}{\frak F}Q{\frak F}^*\Vert= C|\lambda | \,\Vert {\frak F}\partial_tQ{\frak F}^*\Vert =C|\lambda |\,\Vert \partial _{t}Q\Vert.
$$
Finally, 
$$
\Vert \partial _{t}Q(t)\Vert \leq C\Vert \partial _{t}V_{\mathrm{c}}(t)\Vert \leq C\Vert H_{\mathrm{c}}(t)\Vert,
$$
see (\ref{qoft}) and (\ref{vc}). This shows (\ref{m9}).\hfill $\blacksquare $

Lemma \ref{lemmabounds} implies that the sum in (\ref{m10})
converges as $\ell\rightarrow\infty$, and thus 
\begin{eqnarray}
\lefteqn{
\scalprod{\widehat\Psi(\mathbf{1})}{\mathrm{e}^{i\alpha \widehat{{\mathcal{G}}}%
}\widehat\Psi(A)}_{{\widehat{\rm per}}}}\nonumber\\
 &=& \frac{1}{2\pi i}\lim_{\ell \rightarrow \infty }\int_{\gamma _{\ell }}\mathrm{e}^{i\alpha z} \scalprod{\widehat\Psi(\mathbf{1})}{ (z-\widehat{\mathcal{G}}(ir))^{-1}\widehat\Psi(A)}_{{\widehat{\rm per}}} \mathrm{d}z  \notag \\
&&+\sum\limits_{k\in \mathbb{Z}}\left\langle\widehat\Psi(\mathbf{1}),\mathrm{e}^{i\alpha\widehat{{\mathcal{G}}}(ir)}\mathcal{P}_{k}(\lambda )\widehat\Psi(A) \right\rangle_{{\widehat{\rm per}}}  \notag \\
&=&\sum\limits_{k\in \mathbb{Z}}\left\langle \widehat\Psi({\mathbf{1}}),\mathrm{e}^{i\alpha \widehat{{\mathcal{G}}}(ir)}\mathcal{P}_{k}(\lambda )\widehat\Psi(A) \right\rangle_{{\widehat{\rm per}}} +\mathcal{O}(\mathrm{e}^{-\frac{\alpha r}{2}}\lambda ^{2}).  \label{m11}
\end{eqnarray}
To see that the integral above is $\mathcal{O}(\mathrm{e}^{-\frac{\alpha r}{2%
}}\lambda ^{2})$ we observe that by expanding the resolvent in the integrand
in powers of $\lambda$, the terms of order $\lambda^0$ and $\lambda^1$
vanish: the former vanishes since $L_{\mathrm{s}}$ has no spectrum below $\gamma_\ell$, the latter vanishes since the interaction $\widehat{V}_{%
\mathrm{per}}$ is linear in creation and annihilation operators.

The next result shows that the term $k=0$ in (\ref{m11}) is dominant.

\begin{lemma}
\label{lemmam1} We have for all $\alpha \geq 0$ 
\begin{eqnarray*}
&&\left\vert \sum\limits_{\substack{ k\in \mathbb{Z} \\ k\neq 0}}%
\left\langle \widehat\Psi(\mathbf{1}),\mathrm{e}^{i\alpha \widehat{{\mathcal{G}}}(ir)}%
\mathcal{P}_{k}(\lambda )\widehat\Psi(A)\right\rangle_{{\widehat{\rm per}}} \right\vert  \\
&\leq &C\left\Vert A\right\Vert T^{2}\left( 1+\left\Vert \mathrm{e}^{i\alpha 
\widehat{{\mathcal{G}}}_{0}(ir)}\mathcal{P}_{0}(0)-\mathrm{e}^{i\alpha 
\widehat{{\mathcal{G}}}(ir)}\mathcal{P}_{0}(\lambda )\right\Vert \right)
\left( 1+|\lambda |\max_{0\leq t\leq T}\left\Vert H_{\mathrm{c}%
}(t)\right\Vert \right) 
\end{eqnarray*}%
for some $C<\infty $ not depending on $A\in \mathcal{B}(\mathbb{C}^{d})$.
\end{lemma}

\noindent \textit{Proof of Lemma \ref{lemmam1}.\ } Define the family of
unitaries $\{U_{\delta }\}_{\delta \in \mathbb{Z}}$ acting on $\widehat{%
\mathfrak{H}}_{\mathrm{per}}$ by $(U_{\delta }\widehat{\psi})(k):=\widehat{\psi}%
(k+\delta)$. We have $U_{\delta }\widehat{\mathcal{G}}(ir)U_{\delta }^{\ast
}=-\frac{2\pi \delta }{T}+\widehat{\mathcal{G}}(ir)$ and hence 
\begin{equation}
U_{\delta }\mathcal{P}_{k}(\lambda )=\mathcal{P}_{k+\delta }(\lambda
)U_{\delta }.  \label{m13}
\end{equation}
Using this relation we obtain 
\begin{eqnarray*}
\lefteqn{
 \mathrm{e}^{i\alpha \widehat{{\mathcal{G}}}(ir)}\mathcal{P}_{k}(\lambda )\widehat\Psi(A)}\\
&=& \mathrm{e}^{-2\pi i k\alpha/T} U_{k}\mathrm{e}^{i\alpha \widehat{{\mathcal{%
G}}}(ir)}\mathcal{P}_{0}(\lambda )U_{k}^{\ast }\mathcal{P}_{k}(\lambda )\widehat\Psi(A) \\
&=&-\mathrm{e}^{-2\pi i k\alpha/T}U_{k}[\mathrm{e}^{i\alpha \widehat{{\mathcal{G}}}_{0}(ir)}\mathcal{P}_{0}(0)-\mathrm{e}^{i\alpha \widehat{{\mathcal{G}}}(ir)}\mathcal{P}_{0}(\lambda )]U_{k}^{\ast }\mathcal{P}_{k}(\lambda )\widehat \Psi(A)  \\
&&+\mathrm{e}^{-2\pi i k\alpha/T}U_{k}\mathrm{e}^{i\alpha \widehat{{\mathcal{G}%
}}_{0}(ir)}\mathcal{P}_{0}(0)U_{k}^{\ast }\mathcal{P}_{k}(\lambda )\widehat\Psi(A).
\end{eqnarray*}
This equality together with Lemma \ref{lemmabounds} gives the result of
Lemma \ref{lemmam1}. \hfill $\blacksquare$

\medskip 
Combining (\ref{m11}) with Lemma \ref{lemmam1} gives 
\begin{eqnarray*}
\lefteqn{\scalprod{\widehat\Psi(\mathbf{1})}{\mathrm{e}^{i\alpha \widehat{{\mathcal{G}}}}\widehat\Psi(A)}_{{\widehat{\rm per}}}=\scalprod{\widehat\Psi(\mathbf{1})}{\mathrm{e}^{i\alpha \widehat{{\mathcal{G}}}(ir)}\mathcal{P}_{0}(\lambda )\widehat\Psi(A)}_{{\widehat{\rm per}}} } \\
&&\ \ +\Vert A\Vert \left( 1+\left\Vert \mathrm{e}^{i\alpha \widehat{{%
\mathcal{G}}}_{0}(ir)}\mathcal{P}_{0}(0)-\mathrm{e}^{i\alpha \widehat{{%
\mathcal{G}}}(ir)}\mathcal{P}_{0}(\lambda )\right\Vert \right) \,\mathcal{O}\Big( D|\lambda |T+T^{2}+\lambda ^{2}\Big),
\end{eqnarray*}
where $D$ is defined in Theorem \ref{mainthm}. Noticing that $\mathrm{e}%
^{i\alpha \widehat{{\mathcal{G}}}_{0}(ir)}\mathcal{P}_{0}(0)=\mathrm{e}^{i\alpha L_{\mathrm{s}}}\mathcal{P}_{0}(0)$, due to \fer{m13'}, we see that Theorem \ref%
{section extension copy(4)} follows from the following result:

\begin{satz}
\label{thmm2} Suppose the $T$--periodic control term $H_{\mathrm{c}}$
satisfies the dynamical decoupling condition (\ref{dyn.cond}). Then 
\[
\left\Vert \mathrm{e}^{i \alpha L_{\mathrm{s}}}\mathcal{P}_{0}(0)- \mathrm{e}^{i\alpha \widehat{{\mathcal{G}}}(ir)}\mathcal{P}_{0}(\lambda )\right\Vert
\leq C(|\lambda|+\mathrm{e}^{c\alpha|\lambda| T}-1)
\]%
for all $\alpha \geq 0$, where $c,C<\infty $ are constants independent of $\alpha,\lambda,T$ and $H_{\mathrm{c}}$.
\end{satz}

\noindent
\textit{Proof of Theorem \ref{thmm2}.\ } In order to compare $\mathrm{e}^{i\alpha \widehat{{\mathcal{G}}}(ir)}\mathcal{P}_{0}(\lambda )$
with $\mathrm{e}^{i\alpha \widehat{{\mathcal{G}}}_0(ir)}\mathcal{P}_{0}(0)=%
\mathrm{e}^{i \alpha L_{\mathrm{s}}}\mathcal{P}_{0}(0)$, we use Kato's
representation \cite[Chapter I, Section 4.6]{Kato} 
\begin{equation}
\mathcal{P}_{0}(\lambda )=U\mathcal{P}_{0}(0)V,  \label{m20}
\end{equation}
valid for $\lambda$ small enough so that $\left\Vert \mathcal{P}_{0}(\lambda
)-\mathcal{P}_{0}(0)\right\Vert < 1$. The operators $U,V$ are defined as
follows: let $R :=(\mathcal{P}_{0}(\lambda )-\mathcal{P}_{0}(0))^{2}$, then 
\begin{eqnarray*}
U &:=&(1-R)^{-\frac{1}{2}}U^{\prime }=U^{\prime }(1-R)^{-\frac{1}{2}}, \\
V &:=&(1-R)^{-\frac{1}{2}}V^{\prime }=V^{\prime }(1-R)^{-\frac{1}{2}}, \\
U^{\prime } &:=&\mathcal{P}_{0}(\lambda )\mathcal{P}_{0}(0)+(1-\mathcal{P}%
_{0}(\lambda ))(1-\mathcal{P}_{0}(0)), \\
V^{\prime } &:=&\mathcal{P}_{0}(0)\mathcal{P}_{0}(\lambda )+(1-\mathcal{P}%
_{0}(0))(1-\mathcal{P}_{0}(\lambda )).
\end{eqnarray*}
The operator $R$ has norm $\|R\|<1$ and $(1-R)^{-\frac{1}{2}}$ is defined as a bounded operator by the Taylor series of $z\mapsto (1-z)^{-\frac{1}{2}}$, centered at the origin and having radius of convergence one.

Note that $V=U^{-1}$ and that $R$ commutes with both $\cP_0(0)$ and $\cP_0(\lambda)$. Using relation (\ref{m20}) we obtain 
\begin{eqnarray}
\lefteqn{
\exp [i\alpha \widehat{{\mathcal{G}}}(ir)]\mathcal{P}_{0}(\lambda ) =
\mathcal{P}_{0}(\lambda )\exp[i\alpha \widehat{{\mathcal{G}}}(ir)]\mathcal{P}%
_{0}(\lambda)}  \notag \\
&=& U\mathcal{P}_{0}(0)V\mathcal{P}_{0}(\lambda )\exp [i\alpha {\widehat{%
\mathcal{G}}}(ir)]\mathcal{P}_{0}(\lambda)U\mathcal{P}_{0}(0)V  \notag \\
&=& U\mathcal{P}_{0}(0)\exp \left[ i\alpha \mathcal{P}_{0}(0)V\widehat{{%
\mathcal{G}}}(ir)\mathcal{P}_{0}(\lambda )U\mathcal{P}_{0}(0)\right] 
\mathcal{P}_{0}(0)V.\qquad  
\label{m23}
\end{eqnarray}
To understand the last equality, it is important to notice that $U$ maps the range of $\cP_0(0)$ into the range of $\cP_0(\lambda)$ and $V$ maps the range of  $\cP_0(\lambda)$ back into the range of $\cP_0(0)$. Recall that $\widehat{{\mathcal{G}}}_{0}(ir)\mathcal{P}_{0}(0) = L_{\mathrm{s}}\mathcal{P}_{0}(0)$. We prove below that
\begin{equation}
\left\Vert \mathcal{P}_{0}(0)V\widehat{{\mathcal{G}}}(ir)\mathcal{P}_{0}(\lambda )U\mathcal{P} _{0}(0)- L_{\mathrm{s}}\mathcal{P}_{0}(0)\right\Vert \leq c |\lambda | T,  \label{m21}
\end{equation}
for some $c<\infty $ independent of $T, H_{\mathrm{c}}$. Iterating the
Duhamel formula and using the fact that $\exp[i\alpha L_{\mathrm{s}}\mathcal{%
P}_{0}(0)]$ is a contraction group, we obtain from (\ref{m21}) that 
\begin{equation}
\left\|\exp \left[ i\alpha \mathcal{P}_{0}(0)V\widehat{{\mathcal{G}}}(ir)%
\mathcal{P} _{0}(\lambda )U\mathcal{P}_{0}(0)\right] -\exp[i\alpha L_{%
\mathrm{s}}\mathcal{P}_{0}(0)]\right\| \leq \mathrm{e}^{c \alpha |\lambda|
T}-1.  \label{m22}
\end{equation}
Notice that $U,V =\mathbf{1}+ \mathcal{O}(|\lambda|)$, so Theorem \ref{thmm2}
follows from (\ref{m23}) and (\ref{m22}).

It remains to show (\ref{m21}). We will write $\widehat{\mathcal{G}}$ for $%
\widehat{\mathcal{G}}(ir)$ in the remaining part of the proof. Using the
definitions of $U$ and $V$ above and the fact that $R$ commutes with $\cP_0(\lambda)$ and $\cP_0(0)$, we obtain 
\begin{eqnarray}
\lefteqn{\mathcal{P}_{0}(0)V\widehat{{\mathcal{G}}}\mathcal{P}_{0}(\lambda )U\mathcal{P}_{0}(0)}  \notag \\
&=&\mathcal{P}_{0}(0)V^{\prime }(1-R)^{-\frac{1}{2}}\widehat{{\mathcal{G}}}%
\mathcal{P}_{0}(\lambda )(1-R)^{-\frac{1}{2}}U^{\prime }\mathcal{P}_{0}(0) 
\notag \\
&=&\mathcal{P}_{0}(0)\mathcal{P}_{0}(\lambda )(1-R)^{-\frac{1}{2}}\widehat{{%
\mathcal{G}}}\mathcal{P}_{0}(\lambda )(1-R)^{-\frac{1}{2}}\mathcal{P}%
_{0}(\lambda )\mathcal{P}_{0}(0)  \notag \\
&=&\mathcal{P}_{0}(0)(1-R)^{-\frac{1}{2}}\widehat{{\mathcal{G}}}\mathcal{P}%
_{0}(\lambda )(1-R)^{-\frac{1}{2}}\mathcal{P}_{0}(0)  \notag \\
&&+\mathcal{P}_{0}(0)(\mathcal{P}_{0}(\lambda )-\mathcal{P}_{0}(0))(1-R)^{-%
\frac{1}{2}}\widehat{{\mathcal{G}}}\mathcal{P}_{0}(\lambda )(1-R)^{-\frac{1}{%
2}}\mathcal{P}_{0}(0)  \notag \\
&&+\mathcal{P}_{0}(0)\mathcal{P}_{0}(\lambda )(1-R)^{-\frac{1}{2}}\widehat{{%
\mathcal{G}}}\mathcal{P}_{0}(\lambda )(1-R)^{-\frac{1}{2}}(\mathcal{P}%
_{0}(\lambda )\mathcal{P}_{0}(0)-\mathcal{P}_{0}(0)).\qquad  \label{m24}
\end{eqnarray}
The first term on the r.h.s. can be written as 
\begin{eqnarray}
\lefteqn{\mathcal{P}_{0}(0)(1-R)^{-\frac{1}{2}}\widehat{{\mathcal{G}}}%
\mathcal{P}_{0}(\lambda )(1-R)^{-\frac{1}{2}}\mathcal{P}_{0}(0) =\mathcal{P}%
_{0}(0)\widehat{{\mathcal{G}}}\mathcal{P}_{0}(\lambda ) \mathcal{P}_{0}(0)}
\qquad\qquad\qquad\qquad  \notag \\
&&+\mathcal{P}_{0}(0)[(1-R)^{-\frac{1}{2}}-1]\widehat{{\mathcal{G}}}\mathcal{%
P}_{0}(\lambda )\mathcal{P}_{0}(0)  \notag \\
&&+\mathcal{P}_{0}(0)(1-R)^{-\frac{1}{2}}\widehat{{\mathcal{G}}}\mathcal{P}%
_{0}(\lambda )[(1-R)^{-\frac{1}{2}}-1]\mathcal{P}_{0}(0).\qquad  \label{m25}
\end{eqnarray}
By Taylor expanding about $R=0$ and remembering that $R=(\mathcal{P}%
_{0}(\lambda )-\mathcal{P}_{0}(0))^{2}$, we see that $(1-R)^{-\frac{1}{2}}-1
=(\mathcal{P}_{0}(\lambda )-\mathcal{P}_{0}(0))^2\Upsilon (\lambda
)=\Upsilon (\lambda )(\mathcal{P}_{0}(\lambda )-\mathcal{P}_{0}(0))^2$ for
an operator $\Upsilon (\lambda )$ uniformly bounded in norm w.r.t. to $%
\lambda$. It follows from this and (\ref{m24}), (\ref{m25}), together with the fact that $\|\widehat{\cal G}\cP_0(\lambda)\|\leq C$, that 
\begin{eqnarray}
\lefteqn{ \left\| \mathcal{P}_{0}(0)V\widehat{{\mathcal{G}}}\mathcal{P}%
_{0}(\lambda )U \mathcal{P}_{0}(0) - \mathcal{P}_{0}(0)\widehat{{\mathcal{G}}%
} \mathcal{P}_{0}(\lambda ) \mathcal{P}_{0}(0)\right\|}  \label{m26} \\
&&\leq C \Big\{ \| \mathcal{P}_0(0)[\mathcal{P}_0(\lambda) -\mathcal{P}%
_0(0)]\|+ \| [\mathcal{P}_0(\lambda) -\mathcal{P}_0(0)]\mathcal{P}_0(0)\| %
\Big\}  \notag
\end{eqnarray}
for a constant $C<\infty$ independent of $\lambda$, $T$, $H_{\mathrm{c}}$.
Finally, (\ref{m21}) follows from (\ref{m26}) and this result:

\begin{lemma}
\label{lemmmx} For all $T>0$ and any control term $H_{\mathrm{c}}$
satisfying the dynamical decoupling condition (\ref{dyn.cond}) we have 
\[
\left. 
\begin{array}{l}
\left\Vert \mathcal{P}_{0}(\lambda )\mathcal{P}_{0}(0)-\mathcal{P}%
_{0}(0)\right\Vert \\ 
\left\Vert \mathcal{P}_{0}(0)\mathcal{P}_{0}(\lambda )-\mathcal{P}%
_{0}(0)\right\Vert \\ 
\left\Vert \text{ }\widehat{{\mathcal{G}}}(ir)\mathcal{P}_{0}(\lambda )\text{
}\mathcal{P}_{0}(0)-\widehat{{\mathcal{G}}}_{0}(ir)\mathcal{P}%
_{0}(0)\right\Vert%
\end{array}%
\right\} \leq CT|\lambda|
\]%
where $C<\infty $ is a constant not depending on $\lambda, T, H_{\mathrm{c}}$%
.
\end{lemma}

\noindent \textit{Proof of Lemma \ref{lemmmx}. } We start with the first
inequality. We have 
\begin{eqnarray*}
&&\mathcal{P}_{0}(\lambda )\mathcal{P}_{0}(0)-\mathcal{P}_{0}(0) \\
&=&\frac{1}{2\pi i}\oint\limits_{\gamma _{0,\varepsilon }}(\zeta -\widehat{%
\mathcal{G}}(ir))^{-1}\widehat{V}_{\mathrm{per}}\left( ir\right) (\zeta -%
\widehat{\mathcal{G}}_{0}(ir))^{-1}\mathcal{P}_{0}(0)\mathrm{d}\zeta , \\
&=&\frac{1}{2\pi i}\oint\limits_{\gamma _{0,\varepsilon }}\Xi (\zeta )(\zeta -\widehat{%
\mathcal{G}}_{0}(ir))^{-1}\widehat{V}_{\mathrm{per}}\left( ir\right) \mathcal{P}%
_{0}(0)(\zeta -\widehat{\mathcal{G}}_{0}(ir))^{-1}\mathrm{d}\zeta ,
\end{eqnarray*}
where 
\begin{equation}
\Xi (\zeta) = \sum_{n\geq 0}\left[(\zeta-\widehat{\cal G}_0(ir))^{-1} \widehat V_{\rm per}(ir)\right]^n
\label{XI}
\end{equation}
is uniformly bounded for $\zeta \in \gamma_{0,\varepsilon }$ (see before \fer{61.5} for the definition of $\gamma_{0,\varepsilon}$). To establish the first bound in the Lemma it suffices to
show that 
\begin{equation}
\left\Vert (\zeta -\widehat{\mathcal{G}}_{0}(ir))^{-1}\widehat{V}_{\mathrm{per}%
}(ir)\mathcal{P}_{0}(0)\right\Vert \leq C|\lambda |T.  \label{m34'}
\end{equation}%
Let $\varphi _{ij}:=|\varphi _{i}\rangle \langle \varphi _{j}|$, $%
i,j=1,\ldots ,d$, be an orthonormal basis of $\mathfrak{H}_{\mathrm{s}}$
(see before (\ref{m28})). For each $\hat{f}\in \widehat{\mathfrak{H}}_{%
\mathrm{per}}$ and $k\in \mathbb{Z}$ we have 
$$
[\mathcal{P}_{0}(0)\hat{f}](k)=\delta _{k,0}\sum_{i,j}\langle \varphi _{ij}\otimes \Omega _{\mathcal{R}},\hat{f}(0)\rangle _{\mathfrak{H}}\ \varphi _{ij}\otimes \Omega _{\mathcal{R}},
$$ 
where $\delta_{k,0}$ is the Kronecker symbol, and 
\begin{equation}
\lbrack \widehat{V}_{\mathrm{per}}(ir)\varphi _{ij}\otimes \Omega _{\mathcal{R}}](k)=\widehat{v}_{\mathrm{per}}(-k;ir)\varphi _{ij}\otimes \Omega _{\mathcal{R}},
\label{m30}
\end{equation}
see \fer{vhatagain}. We thus have 
\begin{eqnarray}
\lefteqn{\left\Vert  (\zeta -\widehat{\mathcal{G}}_{0}(ir))^{-1}\widehat{V}%
_{\mathrm{per}}(ir)\mathcal{P}_{0}(0)\hat{f} \right\Vert _{\widehat{%
\mathfrak{H}}_{\mathrm{per}}}^{2}}  \notag \\
&\leq &\Vert \hat{f}(0)\Vert _{\mathfrak{H}}^{2}\ \sum_{k\in \mathbb{Z}%
,k\neq 0}\left\Vert \sum_{i,j}(\zeta +\textstyle\frac{2\pi}{T}k-L-ir\widehat N)^{-1}\widehat{v}_{\mathrm{per}}(-k;ir)\,\varphi _{ij}\otimes \Omega _{\mathcal{R}%
}\right\Vert ^{2}\ \ \   \notag \\
&\leq &\Vert \hat{f}\Vert _{\widehat{\mathfrak{H}}_{\mathrm{per}%
}}^{2}d^{2}\max_{i,j}\sum_{k\in \mathbb{Z},k\neq 0}\left\Vert (\zeta +\textstyle\frac{2\pi}{T}k-L-ir)^{-1}\widehat{v}_{\mathrm{per}}(-k;ir)\,\varphi
_{ij}\otimes \Omega _{\mathcal{R}}\right\Vert ^{2}.\ \ \qquad   
\label{m31}
\end{eqnarray}
The sum is only over $k\neq 0$ since we have, by the dynamical decoupling condition \fer{dyn.cond} (see also \fer{twocond}), that $\widehat{Q}(0)=0$, which in turn implies that $\widehat v_{\rm per}(0;ir)=0$, see \fer{vhatagain}.   Again, the number operator is replaced by $\widehat N=1$, since $\widehat{v}_{\mathrm{per}}(-k,ir)\varphi_{ij}\otimes\Omega_\R$ belongs to the one-particle sector. Next we examine the
general term of the sum in (\ref{m31}). The operator $\widehat{v}_{\mathrm{per}%
}(-k;ir)$ has two terms, see \fer{vhatagain}. We deal with the first one, $\lambda\underrightarrow{\widehat Q(-k)}\otimes\Phi_{ir}(g_f)$, the second one is handled in the same way. Writing the
square of the norm in (\ref{m31}) as an inner product, we obtain the
following expression (see \fer{Wtdefo}-\fer{gftheta}) 
\begin{eqnarray}
\lefteqn{
\left\Vert (\zeta +\textstyle\frac{2\pi}{T}k-L-ir)^{-1}\lambda\underrightarrow{\widehat Q(-k)}\otimes\Phi_{ir}(g_f)\,\varphi
_{ij}\otimes \Omega _{\mathcal{R}}\right\Vert ^{2}}\nonumber\\
&=&\frac{\lambda ^{2}}{2}\left\langle \varphi _{ij}\otimes \Omega _{%
\mathcal{R}},\underrightarrow{\widehat{Q}(k)}\otimes a(g_{f,ir})\left\vert
\zeta +\textstyle\frac{2\pi }{T}k-L-ir\right\vert ^{-2}\right.   \notag \\
&&\qquad \qquad \times \left. \underrightarrow{\widehat{Q}(-k)}\otimes
a^{\ast }(g_{f,ir})\ \varphi _{ij}\otimes \Omega _{\mathcal{R}%
}\right\rangle   \notag \\
&=&\max_{e\in \sigma (L_{\mathrm{s}})}\frac{\lambda ^{2}}{2}\int_{\mathbb{R}%
\times S^{2}}|g_{f}(p+ir,\vartheta )|^{2}\left\langle \varphi _{ij},%
\underrightarrow{\widehat{Q}(k)}\left\vert \zeta +\textstyle\frac{2\pi}{T}
k-e-p-ir\right\vert ^{-2}\right.   \notag \\
&&\qquad \qquad \times \left. \underrightarrow{\widehat{Q}(-k)}\ \varphi
_{ij}\right\rangle \d p\,\d \vartheta .  \notag
\end{eqnarray}
Recall the notation \fer{gftheta}. 
We have used the standard \textquotedblleft pull-through\textquotedblright\
method to arrive at the last expression (e.g., $a(p,\vartheta)L_\R=(L_\R+p)a(p,\vartheta)$). With $\Vert \underrightarrow{%
\widehat{Q}(k)}\Vert \leq \Vert Q\Vert $ we obtain 
\begin{eqnarray}
\lefteqn{\left\Vert (\zeta -\widehat{\mathcal{G}}_{0}(ir))^{-1}\widehat{V}%
_{\mathrm{per}}(ir)\mathcal{P}_{0}(0)\hat{f} \right\Vert _{\widehat{%
\mathfrak{H}}_{\mathrm{per}}}^{2}}  \label{m32} \\
&\leq &\lambda ^{2} d^{2}\Vert \hat{f}\Vert _{\widehat{\mathfrak{H}}_{\mathrm{per}%
}}^{2}\Vert Q\Vert ^{2}\sum_{k\in \mathbb{Z},k\neq 0}\max_{e\in
\sigma (L_{\mathrm{s}})}\int_{\mathbb{R}\times S^{2}}\left\vert \frac{%
g_{f}(p+ir,\vartheta )}{\zeta +\frac{2\pi}{T}k-e-p-ir}\right\vert
^{2}\d p\,\d \vartheta .  \notag
\end{eqnarray}
Set
$$
h(u)=\int_{S^2} |g_f(u-e+{\rm Re}\zeta+ ir,\vartheta)|^2\d \vartheta,\qquad M=\frac{2\pi}{T}k, \qquad a=(r-{\rm Im}\zeta)^2.
$$
Then
\begin{equation}
\int_{\mathbb{R}\times S^{2}}\left\vert \frac{ g_{f}(p+ir,\vartheta )}{\zeta +\frac{2\pi}{T}k-e-p-ir}\right\vert
^{2}\d p\,\d \vartheta = \int_{\mathbb R} \frac{h(u)}{(u-M)^2+a^2}\d u.
\label{m32.1}
\end{equation}
We have 
$$
\sup_{M\in{\mathbb R}}\frac{M^2}{(u-M)^2+a^2} = \frac{(u^2+a^2)^2}{a^4+u^2a^2}
$$
and therefore
$$
\int_{\mathbb R} \frac{h(u)}{(u-M)^2+a^2}\d u \leq M^{-2}\int_{{\mathbb R}} \frac{(u^2+a^2)^2}{a^4+u^2a^2}h(u)\d u\leq CM^{-2}= C\left(\frac{T}{2\pi k}\right)^2,
$$
due to condition (A2), equation \fer{m34}. Combining this with \fer{m31}, \fer{m32} and \fer{m32.1}, we obtain \fer{m34'}.

The second bound of Lemma \ref{lemmmx}, $\|\mathcal{P}_0(0)\mathcal{P}%
_0(\lambda)-\mathcal{P}_0(0)\|\leq CT|\lambda|$ is established analogously.

To show the last bound in Lemma \ref{lemmmx}, write (see also \fer{XI})
\begin{eqnarray*}
\lefteqn{\widehat{{\mathcal{G}}}(ir)\mathcal{P}_{0}(\lambda )\text{ }\mathcal{P}%
_{0}(0)-\widehat{{\mathcal{G}}}_{0}(ir)\mathcal{P}_{0}(0) =\mathcal{P}_{0}(\lambda )\widehat{{\mathcal{G}}}(ir)\text{ }\mathcal{P}%
_{0}(0)-\widehat{{\mathcal{G}}}_{0}(ir)\mathcal{P}_{0}(0)} \\
&=&\mathcal{P}_{0}(\lambda )(\widehat{V}_{\mathrm{per}}(ir)+L_{\mathrm{s}})%
\mathcal{P}_{0}(0)-\mathcal{P}_{0}(0)L_{\mathrm{s}}\mathcal{P}_{0}(0) \\
&=&\frac{1}{2\pi i}\oint\limits_{\gamma _{0,\varepsilon }}\Xi (\zeta )(\zeta -\widehat{%
\mathcal{G}}_{0}(ir))^{-1}\widehat{V}_{\mathrm{per}}\left( ir\right) \mathcal{P}%
_{0}(0)\mathrm{d}\zeta  \\
&&+\frac{1}{2\pi i}\oint\limits_{\gamma _{0,\varepsilon }}\Xi (\zeta )(\zeta -\widehat{%
\mathcal{G}}_{0}(ir))^{-1}\widehat{V}_{\mathrm{per}}\left( ir\right) \mathcal{P}%
_{0}(0)(\zeta -\widehat{\mathcal{G}}_{0}(ir))^{-1}L_{\mathrm{s}}\mathcal{P}%
_{0}(0)\mathrm{d}\zeta .
\end{eqnarray*}%
By the bound (\ref{m34'}) on $(\zeta -\widehat{\mathcal{G}}_{0}(ir))^{-1}%
\widehat{V}_{\mathrm{per}}\left( ir\right) \mathcal{P}_{0}(0)$, the third
inequality in Lemma \ref{lemmmx} holds. This completes the proof of Lemma \ref{lemmmx} and with that the proof of
Theorem \ref{thmm2}. \hfill $\blacksquare$

\end{document}